\documentclass[%
 reprint,
 amsmath,amssymb,
 aps,
]{revtex4-2}

\usepackage{comment}
\usepackage{graphicx}
\usepackage{dcolumn}
\usepackage{bm}
\usepackage{hyperref}
\usepackage[dvipsnames]{xcolor}
\usepackage{cancel}

\def\<{\langle}
\def\>{\rangle}
\def\{{\lbrace}
\def\}{\rbrace}
\def\({\left(}
\def\){\right)}
\def\beq{\begin{equation}}
\def\eeq{\end{equation}}

\newcommand{\av}[1]{\left\langle #1 \right\rangle}

\newcommand{\ext}{\text{ext}}

\newcommand{\info}{\text{info}}
\newcommand{\sto}{\text{stop}}
\newcommand{\sts}{\text{s}}
\newcommand{\inn}{\text{in}}
\newcommand{\out}{\text{\ext}}
\newcommand{\calX}{\mathcal{X}}

\newcommand{\eff}{\text{eff}}

\newcommand{\simm}{\text{sim}}
\newcommand{\tot}{\text{tot}}
\newcommand{\asy}{\text{asy}}

\newcommand{\Hseq}{H_{\sts}}  
\newcommand{\dHs}{H'_{\sts}}
\newcommand{\dSinfoPM}{\Delta\tilde{S}_\info}

\newcommand{\bmCk}{\bm{C}_1^k}
\newcommand{\bmCkant}{\bm{C}_1^{k-1}}

\begin{document}

\preprint{APS/123-QED}

\title{Information in feedback ratchets}

\author{Natalia Ruiz-Pino$^{1,2}$}
\author{Daniel Villarubia-Moreno$^{2,3}$}
\author{Antonio Prados$^{1}$}
 \email{prados@us.es}  
\author{Francisco J. Cao-García$^{2,4}$}
 \email{francao@ucm.es}  
  \affiliation{$^{1}$ Dpto. Física Teórica, Apartado de Correos 1065, Universidad de Sevilla, E-41080 Sevilla, Spain.}
  \affiliation{$^{2}$ Dpto. Estructura de la Materia, Física Térmica y Electrónica, Universidad Complutense de Madrid, Plaza de Ciencias, 1, 28040 Madrid, Spain.}
  \affiliation{$^{3}$ Dpto. de Matemáticas (GISC), Universidad Carlos III de Madrid, Spain.}
  \affiliation{$^{4}$ Instituto Madrileño de Estudios Avanzados en Nanociencia, IMDEA Nanociencia, Calle Faraday, 9, 28049 Madrid, Spain.}

\date{\today}

\begin{abstract}
Feedback control uses the state information of the system to actuate on it. The information used implies an effective entropy reduction of the controlled system, potentially increasing its performance. How to compute this entropy reduction has been  formally  shown for a general system, and has been explicitly computed for spatially discrete systems. Here, we address a relevant example of how to compute the entropy reduction by information in a spatially continuous feedback-controlled system. Specifically, we consider a feedback flashing ratchet, which constitutes a paradigmatic example for the role of information and feedback in the dynamics and thermodynamics of transport induced by the rectification of Brownian motion. A Brownian particle moves in a periodic potential that is switched on and off by a controller, with the latter performing the switching depending on the system state. We show how the entropy reduction can be computed from the entropy of a sequence of control actions, and also discuss the required sampling effort for its accurate computation. Moreover, the output power developed by the particle against an external force is investigated, which---for some values of the system parameters---is shown to become larger than the input power due to the switching of the potential: the apparent efficiency of the ratchet thus becomes higher than one, if the entropy reduction contribution is not considered. This result highlights the relevance of including the entropy reduction by information in the thermodynamic balance of feedback controlled devices, specifically when writing the second principle: the inclusion of the entropy reduction by information leads to a well-behaved efficiency over all the range of parameters investigated.
\end{abstract}

\maketitle


\section{INTRODUCTION \label{Sec:Introduction}}

Information on the system state can be used to increase its performance. Feedback control, which is an old engineering topic~\cite{bechhoefer_feedback_2005}, employs information in real-time to optimize the control actuation on the system. Feedback control has also been investigated from the physical standpoint,  especially in relation to the Maxwell demon and the Szilard engine, which seemingly challenged thermodynamic principles~\cite{Leff2002,Szilard1929,Szilard1964}. After the development of Information Theory~\cite{Shannon1949,cover_elements_2006}, these problems have also attracted the attention of the Computation Theory community, leading to the formulation of the Landauer principle~\cite{Landauer1961,Bennett1982}. 

More recently, feedback control and the thermodynamic role of information has been an active field of research in statistical physics from various perspectives~\cite{Touchette2000,Touchette2004,Quan2006,Sagawa2008,cao_information_2009,cao_thermodynamics_2009,Mandal2012,Barato2014b,van_vu_uncertainty_2020,lozano-duran_information-theoretic_2022,bhattacharyya_feedback-controlled_2022}. Among other things, this interest has been motivated by the increase of performance observed in stochastic systems under feedback control~\cite{Touchette2000,Touchette2004,cao_feedback_2004,cao_information_2009,abreu_thermodynamics_2012,lucero_maximal_2021}. This has also lead to experimental realizations of these systems, where information is used through feedback to correct the stochastic fluctuations~\cite{Lopez2008,Toyabe2010,Debiossac2020,lucero_maximal_2021,saha_bayesian_2022}. The main aim is to understand the limits of performance of physical and biological systems under feedback control~\cite{feito_optimal_2009,Barato2014h,Barato2015,lucero_maximal_2021,saha_bayesian_2022,lee_speed_2022}, and construct the extensions of thermodynamics and out-of-equilibrium statistical physics required to study these systems~\cite{Sagawa2008,cao_information_2009,abreu_thermodynamics_2012,seifert_stochastic_2012,Sagawa2013a,Bechhoefer2015,Potts2018,Potts2019,van_vu_uncertainty_2020}.

Brownian ratchets or Brownian engines are devices that generate a particle flow by rectifying Brownian  motion~\cite{astumian_fluctuation_1994,Sekimoto1997,reimann_introduction_2002,astumian_brownian_2002}. Flashing ratchets are a type of Brownian ratchets, which perform the rectification by switching on/off  a periodic asymmetric potential acting on the particle~\cite{astumian_fluctuation_1994,Reimann2002,cao_feedback_2004,feito_time-delayed_2007,craig_effect_2008,Lopez2008,feito_optimal_2009,roca_optimal_2014}. This alternation, between the motion under the action of the potential---while it is on---and free diffusion---while it is off, rectifies the Brownian motion and may create a flow in a given direction.  The switching (on or off) of the potential can be controlled by different mechanisms: chemical reactions, thermal fluctuations, randomness, etc. These switching controls can be classified into two main  categories: (i) open-loop controls, which operate without collecting information on the state of the system, and (ii) feedback- or closed-loop controls, in which the on or off switching depends on the information obtained by the control on the system state. 

\begin{figure}
    \centering    
    \includegraphics[width=0.6\linewidth]{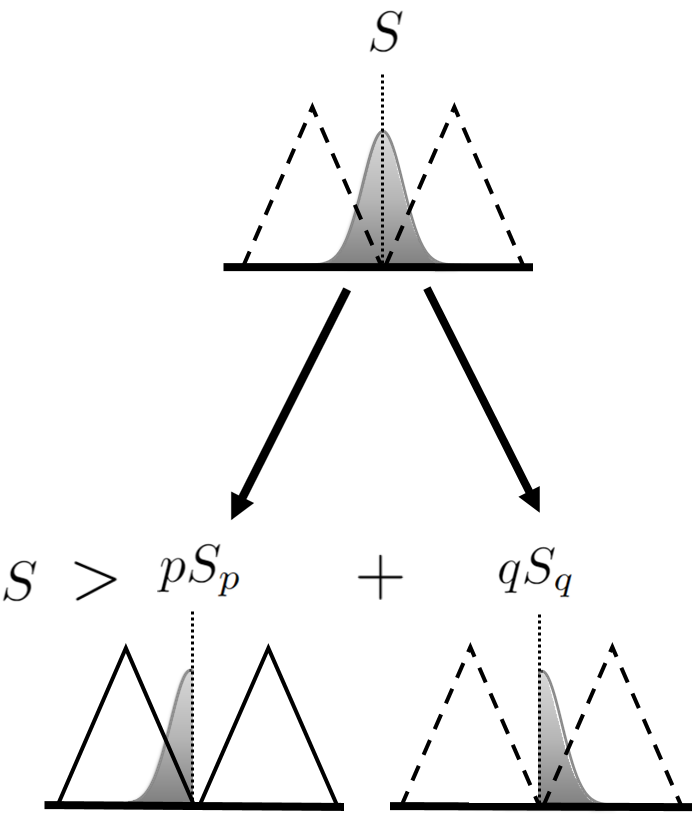}
    \caption{Entropy reduction by information gathered by the feedback ratchet controller. Additional information on the state of the system further determines the macrostate by reducing the number of compatible microstates~\cite{cao_information_2009} and, thus, also reduces entropy. In the top panel, the probability distribution of the particle microstates prior to the measurement is plotted, this macrostate has entropy $S$. The measurement has two possible outputs: the particle is found at the left (of the vertical dashed line), with probability $p$ and associated entropy $S_p$, or at the right, with probability $q = 1-p$ and associated entropy $S_q$.  The additional information concentrates the probability in the microstates compatible with the result of the measurement, reducing the entropy of the resulting macrostate. Therefore, the average entropy after the measurement $p S_p + q S_q$ is lower than the entropy before the measurement $S$ [See Eq.~\eqref{eq:Sp-def}]. 
    }
    \label{fig:entropia}
\end{figure}

Feedback flashing ratchets provide simple models to understand the performance increase due to the information used in feedback control. The study of the role of information in feedback ratchets~\cite{cao_information_2009,Feito2007,abreu_thermodynamics_2012} have led to the development of a general theory for the role played by information in the dynamics and thermodynamics of feedback-controlled systems \cite{cao_thermodynamics_2009}. This general theory shows that the information gathered by the controller on the state of the system reduces the number of microstates compatible with the macrostate of the system. It is the associated reduction of the entropy of feedback-controlled systems, as qualitatively depicted in Fig.~\ref{fig:entropia}, that allows them to outperform their open-loop counterparts~\cite{cao_feedback_2004,cao_information_2009}. 

One of the main goals of this paper is to compute the efficiency and the entropy reduction rate by information in spatially continuous feedback flashing ratchets, applying the general theory developed in Ref.~\cite{cao_thermodynamics_2009}. This work extends to the spatially continuous case the results found for spatially discrete feedback flashing ratchets in Ref.~\cite{jarillo_efficiency_2016}. Entropy reduction by information has been shown to increase both the flux~\cite{cao_feedback_2004} and the usable extractable work~\cite{cao_thermodynamics_2009}. The entropy reduction is given by the entropy rate of control actions, which takes into account that no additional entropy reduction---or increase of performance---is obtained with redundant information~\cite{cao_thermodynamics_2009}. The procedures developed here to compute the information rate and the entropy reduction from numerically obtained control sequences can also be applied to obtain these magnitudes from experimental control sequences.

In this work, we also present an alternative, physically appealing, derivation for the entropy reduction of the feedback controlled ratchet. Instead of considering the  controller as an external agent acting on the system, as in Ref.~\cite{cao_thermodynamics_2009}, we show that the entropy reduction by information can also be obtained by explicitly writing down the evolution of the joint probability distribution of the particle position and the control history at the time instants when the control is updated. The using without error of the position of the particle by the controller entails that the total entropy, i.e., that of the particle plus the controller, is continuous at the control updates. In turn, this entails that the entropy of the particle is reduced thereat, since the entropy of the control monotonically increases at each control update~\cite{cover_elements_2006,cao_thermodynamics_2009}.

Our paper is organized as follows, in Sec.~\ref{sec:model}, we present the feedback controlled ratchet, describing in detail both its time evolution between control updates and how the control is updated at regular times. Section~\ref{sec:entropy-def} is devoted to the analysis of the entropy of the feedback controlled ratchet. First, in Sec.~\ref{sec:definitions}, we put forward the basic definitions. Second, in Sec.~\ref{sec:long-time-entropy}, we study the long-time behavior of the system and discuss the entropy reduction due to the information gathered by the controller. The thermodynamic balance is investigated in Sec.~\ref{sec:thermodyn-balance}. By using the first and second principles, we show the relevance of the entropy reduction in the definition of the efficiency of the ratchet. The methods employed to obtain the numerical results are described in depth in Sec.~\ref{sec:numerical-methods}. The results are presented in Sec.~\ref{sec:results}, where we show (i) the system reaching a well-defined long-time regime, (ii) the dependence of the entropy reduction on the parameters of our model: control update time $\Delta t_m$, asymmetry of the potential $a$, potential height $V_0$, and external force $F_{\ext}$, and (iii) the input/output power and the efficiency as a function of $F_{\ext}$. Finally, a discussion of the results of the paper, together with the main conclusions and perspectives for future work, is provided in Sec.~\ref{sec:conclusions}. The Appendices deal with some technicalities that are omitted in the main text.

\section{MODEL: FEEDBACK FLASHING RATCHET}\label{sec:model} 

\begin{figure}
    \centering
    \includegraphics[width=0.70\linewidth]{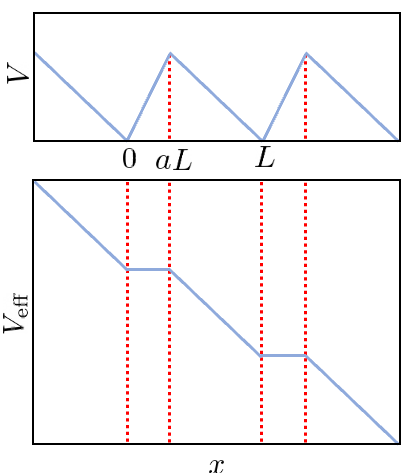}
    \caption{Potential (top panel) and effective potential emerging in the limit $\Delta t_m\to 0$ (bottom panel), as a function of the position of the particle. In the limit $ \Delta t_m \to 0^+$, the control acts continuously and thus the potential $V(x)$ is switched on (off) for positions giving a positive (negative) force $-V'(x)$,  $-V'(x)>0$ ($-V'(x)<0$). This yields the effective potential $V_{\eff}$ shown in the bottom panel.
    }  
    \label{fig:potencial-efectivo}
\end{figure}

The Brownian engine chosen for this study is the single-particle feedback flashing ratchet~\cite{Reimann2002,reimann_introduction_2002,astumian_brownian_2002}. Our Brownian particle moves on a line under the action of 
a periodic potential $V(x)$. The feedback control measures the position of the particle at regular time intervals $\Delta t_m$, and updates the potential acting on the particle depending on the instantaneous position of the particle. More specifically, at time $t_k=k\Delta t_m$, the feedback controller measures $x(t_k)$ and decides whether the potential is on or off in the time interval $J_k\equiv(t_k,t_{k+1})$: If the associated force $-V'(x(t_k))$ is positive, the potential is turned on; while if the force is negative, the potential is turned off. Physically, the action of this feedback controller creates a flux to the right of the particle, see Fig.~\ref{fig:potencial-efectivo} for a qualitative picture. Mathematically, the action of the controller can be thus characterized by a dichotomic variable $C_k$, equal to either $0$ (potential off) or $1$ (potential on): the potential acting on the particle in the time interval $J_k$ is $C_k V(x)$. In addition, there is an external force $-F_{\ext}$, against the induced particle flux.

Here, we focus on the overdamped regime, in which the velocity degree of freedom of the particle instantaneously relaxes to equilibrium at temperature $T$. Therefore, in the time interval $J_k$ between control updates, the position of the Brownian particle obeys the Langevin equation
\beq
\gamma\dot{x} = -C_k\, V'(x)-F_{\ext}+\xi(t),
\label{Eq:Langevin2}
\eeq
where $\gamma$ is the friction coefficient and $\xi(t)$ represents the effect of thermal fluctuations, modeled by a  Gaussian white noise:
\beq
\langle \xi(t)\rangle=0, \quad \langle \xi(t)\xi(t')\rangle = 2 k_B T \gamma \delta(t-t'),
\label{Eq:Noise}
\eeq
with $k_B$ being the Boltzmann constant. 

For the potential $V(x)$, we choose a $L$-periodic piecewise linear potential, with height $V_0$ and asymmetry $a$:
\beq
V(x) = \left\{ \begin{array}{lcc}
		\frac{V_0}{a }\frac{x}{L}, & \hspace{0.2cm} \text{if} \hspace{0.2cm}	 \text{mod}_1(x/L) \leq a, \\
		\\
		\frac{V_0}{1-a}\left(1-\frac{x}{L}\right), & \hspace{0.2cm} \text{if} \hspace{0.2cm}	\text{mod}_1(x/L) > a,	\\
		\end{array}
		\right.
\label{Eq:Potential}
\eeq
where $\text{mod}_1()$ provides the decimal part of its argument. This switchable potential gives a force
\beq
-V'(x) = \left\{ \begin{array}{lcc}
		-\frac{V_0}{a L}<0, & \hspace{0.2cm} \text{if} \hspace{0.2cm}	 \text{mod}_1(x/L) \leq a, \\
		\\
		\frac{V_0}{(1-a)L}>0, & \hspace{0.2cm} \text{if} \hspace{0.2cm}	\text{mod}_1(x/L) > a,	\\
		\end{array}
		\right.
\label{Eq:force}
\eeq

The controller measures the particle position at $t_k=k\Delta t_m$, and tries to maximize the force to the right: thus it switches on (off) the potential in the time interval $J_k=(t_k,t_{k+1})$ if the force $-V'(x)$ at the particle position $x(t_k)$ is positive (negative). Therefore, we have
\beq
C_k = \left\{ \begin{array}{lcc}
		0, & \hspace{0.2cm} \text{if} \hspace{0.2cm}	 \text{mod}_1(x(t_k)/L) \leq a, \\
		\\
		1, & \hspace{0.2cm} \text{if} \hspace{0.2cm}	\text{mod}_1(x(t_k)/L) > a.	\\
		\end{array}
		\right.
\label{Eq:alpha}
\eeq
The value $C_k$ is kept during the whole  time interval $J_k$, until the next control update at time $t_{k+1}$.
This \textit{maximization protocol} promotes the actuation of the part of the potential providing positive forces, thus giving a net particle flux to the right---see discussion below.  

The transport induced by the on-off switching is known as the ratchet effect \cite{astumian_fluctuation_1994, Reimann2002}. Open-loop controlled ratchets, i.e., without feedback of the particle position, require an asymmetric potential, $a\ne 1/2$, to induce such a transport. For example, if the controller switches on-off the potential periodically, a symmetric potential, $a=1/2$, gives a vanishing flux in average, because a given trajectory and its mirror reflection are equally probable \cite{Reimann2002}. 

The introduction of feedback enhances the ratchet effect and facilitates the emergence of flux even for symmetric potentials. It is especially illuminating to consider the limiting case $\Delta t_m \to 0$, in which the feedback controller knows the position of the particle at all times. This instant maximization protocol has been proven to give the maximum power for the single-particle feedback flashing ratchet~\cite{cao_feedback_2004,feito_time-delayed_2007,craig_effect_2008,feito_optimal_2009}. 
Therein, the potential is always on (off) for positions such that the force $-V'$ is positive (negative), which leads to an effectively asymmetric potential $V_{\eff}$---see below and also Ref. \cite{cao_feedback_2004}.  

The above physical picture is illustrated by Fig.~\ref{fig:potencial-efectivo}, which shows the potential $V(x)$ (top panel) and the effective potential $V_{\eff}(x)$ (bottom panel). The Brownian particle is pushed to the right where $-V_{\eff}'>0$ (in this region, $-V'_{\eff}=-V'$), whereas it freely diffuses where $V_{\eff}'=0$ (in this region, the original force $-V'<0$). A directed flux to the right stems from the combination of (i) the confinement to the minima of the original potential due to the decreasing parts of $V_{\eff}$ and (ii) the free diffusion in the flat parts of $V_{\eff}$, which facilitates reaching the next minimum to the right.  This flow persists even against an opposing external force $F_{\ext}$, in a range of values  $0\le F_{\ext}\le F_{\sto}$; it vanishes for  $F_{\ext}=F_{\sto}$.

\section{Entropy of the feedback controlled
  ratchet}\label{sec:entropy-def}

\begin{figure}
    \centering
    \includegraphics[width=1.0\linewidth]{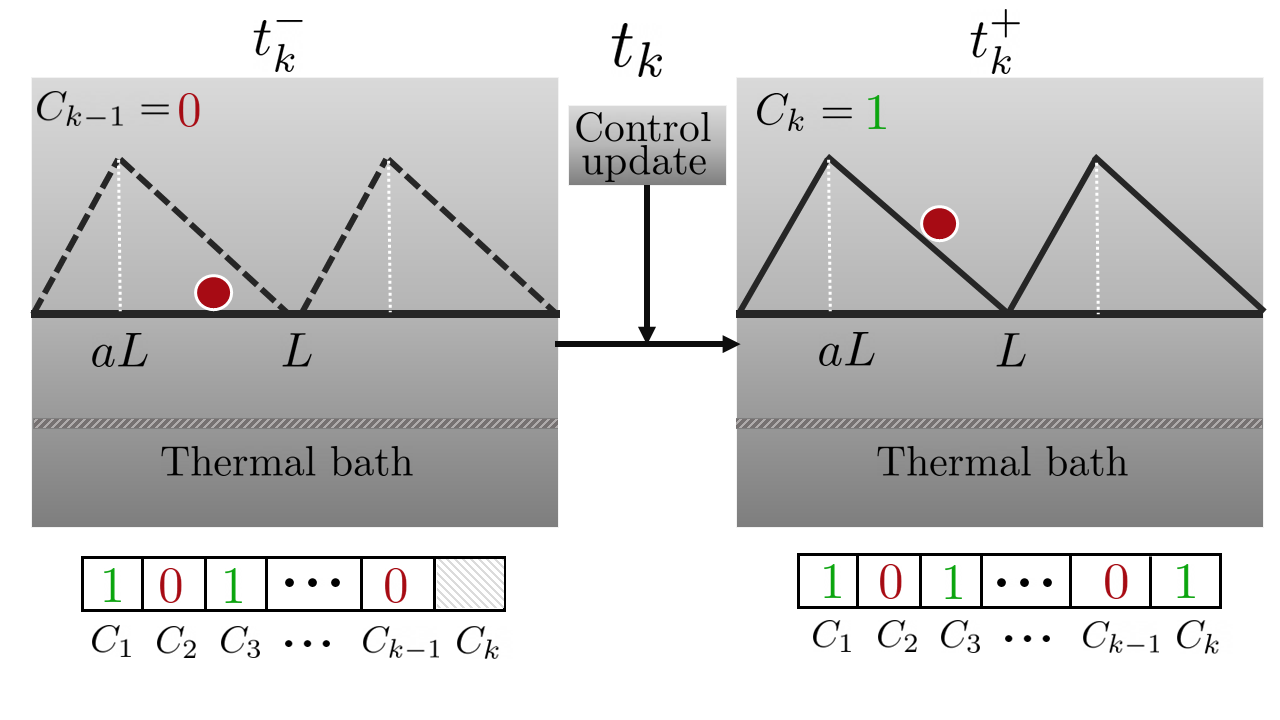}
    \caption{
    Conceptual scheme of the feedback ratchet for the one-particle system considered in this work. The Brownian particle moves in a periodic potential $V(x)$, given by Eq.~\eqref{Eq:Potential}, that can be either off (dashed lines) or on (solid lines), and it is also coupled to a thermal bath at temperature $T$. A feedback controller measures the state of the system  at time $t_k$, i.e. $x(t_k)$, and sets the next bit of the controller memory ($C_k = 0, 1$) accordingly, following Eq.~\eqref{Eq:alpha}, so that the potential acting on the particle for $t=t_k^+$ is $C_k V(x)$. In the specific trajectory portrayed here, the potential was off at time $t_k^-$ and, since $x(t_k^-)$ lies on an interval where the force $-V'$ is positive, the potential is switched on for $t=t_k^+$.}
    \label{fig:esquema}
\end{figure}

Here, we put forward a physically motivated discussion of the entropy of the feedback controlled ratchet, and the associated entropy reduction by information. The controller is assumed to have a---previously reset---memory where it stores the control actions, i.e., it contains the minimal information used by the controller. See Fig.~\ref{fig:esquema} for a sketch of the system. We show below that the entropy reduction of the particle is in this case equal to the increase of entropy of the control memory, as we consider an error-free deterministic control.  The result obtained for the entropy reduction of the particle coincides with the maximum reduction predicted by the more general approach of Ref.~\cite{cao_thermodynamics_2009}, which considered the control as an external agent acting on the particle system---without assumptions on its internal structure. 

\subsection{Definitions}\label{sec:definitions}

The regular update of the controller naturally divides the time axis into time windows of width $\Delta t_{m}$. We have already labeled these time windows as $J_{k}\equiv (t_{k},t_{k+1})$, i.e., by the number of updates of the controller $k$ prior to the time interval $J_{k}$. Throughout each interval $J_k$, the control takes a value $C_k$: $C_k=1$ ($C_k=0$) if the potential is switched on (off) in the $k$-th interval $J_{k}$. Let $\bmCk$ be the vector with the history of control actions from its more recent $k$-th value to the more ancient first value, 
\begin{equation}
    \bmCk\equiv\{C_{k},C_{k-1}\ldots,C_{2},C_{1}\}.
    \label{eq:bmCk-def}
\end{equation}

The entropy of the particle plus the controller in the $k$-th time interval $J_{k}$ is defined as
\begin{equation}
  \label{eq:Spc-kth-window}
  S^{(k)}(t)=-k_{B}\sum_{\bmCk}\int dx\, P(x,\bmCk,t)\ln
  P(x,\bmCk,t) .
\end{equation}
Now we make use of the Bayes theorem $P(x,\bmCk;t)=P(x|\bmCk;t)P(\bmCk)$, where $P(\bmCk)$ does not depend on $t$ because the state of
the control does not change in $J_{k}$---the state of the controller
is only updated at the set of discrete points $t_{k}$~\footnote{
Control memory positions $C_i$ with $i>k$ do not contribute to the entropy until they store a control action, as all or them were initially reset to a fixed value, e.g., $C_i=0$. 
}.
Repeated use of the Bayes theorem in Eq.~\eqref{eq:Spc-kth-window}  gives
\begin{align} 
  S^{(k)}(t)=& -k_{B} \sum_{\bmCk} P(\bmCk)\ln
    P(\bmCk)\nonumber \\
    &- k_{B}\sum_{\bmCk} P(\bmCk) \int dx\,
    P(x|\bmCk;t)  \ln P(x|\bmCk;t).
     \label{eq:Spc-kth-window-v2}
\end{align}
Identifying the first term on the right hand side (rhs) with the entropy
of the controller in the time interval $J_k$, the second term is the entropy of the
particle, i.e.,
\begin{subequations}\label{eq:Sc-and-Sp}
  \begin{align}
    S^{(k)}(t)& =S_{c}^{(k)}+S_{p}^{(k)}(t), \\
    S_{c}^{(k)}&=-k_{B} \sum_{\bmCk} P(\bmCk)\ln
                 P(\bmCk), \label{eq:Sc-def} \\
    S_{p}^{(k)}(t)&=-k_{B}\sum_{\bmCk} P(\bmCk) \int dx\,
                    P(x|\bmCk;t)  \ln P(x|\bmCk;t).
                    \label{eq:Sp-def}
\end{align}
\end{subequations}
Note that Eq.~\eqref{eq:Sp-def} is precisely the result for the entropy
of the particle after $k$ measurements obtained in Ref.~\cite{cao_thermodynamics_2009}, see
Eqs.~(3) and (5) therein.

Let us consider the entropy balance at any of the times $t_{k}$ at
which the feedback controller is updated. On intuitive grounds, since the
deterministic feedback controller has a complete information about the
particle state, as given by its position $x$, and its value is updated without any uncertainty, the entropy of the system plus the
controller $S$ should be continuous at $t_{k}$: $S^{(k-1)}(t_{k}^{-})=S^{(k)}(t_{k}^{+})$, i.e.,
\begin{equation}\label{eq:S-cont-at-tk}
  S_{p}^{(k-1)}(t_{k}^{-})+S_{c}^{(k-1)}=S_{p}^{(k)}(t_{k}^{+})+S_{c}^{(k)}, 
\end{equation}
see Appendix~\ref{sec:app-S-cont} for a rigorous justification. We
have brought to bear that the control has acted $k-1$ times at
$t=t_{k}^{-}$.  
We can thus calculate the change of entropy of the
particle due to the $k$-th update of the control:
\begin{equation}
  \label{eq:deltaS-p-k}
  \Delta S_{pc}^{(k)}\equiv S_{p}^{(k)}(t_{k}^{+})-S_{p}^{(k-1)}(t_{k}^{-})=-\Delta
  S_{c}^{(k)},
\end{equation}
where
\begin{equation}
  \label{eq:deltaS-p-k-v2}
  \Delta S_{c}^{(k)}\equiv S_{c}^{(k)}-S_{c}^{(k-1)}.
\end{equation}
Making use of the Bayes theorem for the control history, in the
form $P(\bmCk)=P(C_{k}|\bmCkant)P(\bmCkant)$, and of Eq.~\eqref{eq:Sc-def}, we have
\begin{align}
  \Delta S_{c}^{(k)}
   =-k_{B}\sum_{\bmCk}P(\bmCk)\ln
     P(C_{k}|\bmCkant)\ge 0.
     \label{eq:deltaS-p-k-v3}
\end{align}
Since all the terms in the sum are non-negative, the equality only
holds if all terms vanish. For each term to vanish, there are two
possibilities: (i) $P(C_{k}|\bmCkant)=0$, which entails 
$P(\bmCk)=0$ applying Bayes's theorem, or (ii)
$P(C_{k}|\bmCkant)=1$. This means that the only possibility for
having $\Delta S_{c}^{(k)}=0$ is to have a deterministic $C_{k}$ for a
given previous history of control actions $\bmCkant$.

\subsection{Long-time regime. Entropy reduction}\label{sec:long-time-entropy}

The joint action of the controller and the external force leads the system to a long-time non-equilibrium regime, in which the particle moves---in average---with a constant average velocity $v_{\sts}$, as long as the external force is smaller than a certain maximum value $F_{\sto}$, i.e., $F_{\ext}<F_{\sto}$. This long-time non-equilibrium regime does not lead to a regular steady state, since the control is updated at regular time intervals of length $\Delta t_{m}$. We expect the long-time regime to lead to  ensemble averaged magnitudes with period $\Delta t_{m}$~\footnote{Strictly speaking, we expect a time-periodic long-time behavior for the $x$ modulo $L$ problem, i.e., with periodic boundary conditions.}. See Sec.~\ref{sec:results-long-time} for a numerical check of this expectation.

In this long-time regime, certain expressions are simplified, as shown below. We consider the long time limit $t>t_{r}$, where $t_{r}$ is a characteristic relaxation time to reach the long-time behavior. Consistently, the number of updates of the control has been also large, $k>k_{r}=t_{r}/\Delta t_{m}$, with $k_{r}\gg 1$. Below, we employ the subindex $s$ to remark that the physical quantities are evaluated in this long time regime.

Let us assume the following property, which we will later check numerically in our system: There exists a certain integer $M_{0}$ such that the conditioning in Eq.~\eqref{eq:deltaS-p-k-v3} can be (approximately) reduced to only the $M$ closer control actions for long enough times, i.e., large enough $k$,
\begin{equation}\label{eq:M0-def}
P_{\sts}(C_{k}|\bm{C}^{k-1}_{k-M}) = P_{\sts}(C_{k}|\bm{C}^{k-1}_{k-M_0} 
), \quad \forall M\ge M_{0},
\end{equation}
where $\bm{C}^{k-1}_{k-M} \equiv (C_{k-1},C_{k-2},\ldots,C_{k-M})$, consistently with Eq.~\eqref{eq:bmCk-def}.    
In other words, the influence of the history of
controller actions is restricted to the previous $M_{0}$ values; the
``older'' values become irrelevant for long times. From
Eqs.~\eqref{eq:deltaS-p-k-v3} and \eqref{eq:M0-def}, the entropy change of the control
$\Delta S_{c,s}^{(k)}$ in the long-time limit is
\begin{align}
 \Delta S_{c,s}^{(k)} =-k_{B}\!\sum_{C_{k}}\!\sum_{C_{k-1}} \cdots
  \!\sum_{C_{k-M}}  P_{\sts}(& \bm{C}^k_{k-M})
 \ln P_{\sts}(C_{k}|\bm{C}^{k-1}_{k-M}), \nonumber \\
& \forall M\ge M_0.
    \label{eq:deltaSc-k-indep-k}
\end{align}
In the terminology of information theory~\cite{cover_elements_2006}, the rhs of Eq.~\eqref{eq:deltaSc-k-indep-k} is known as the entropy rate---here, of the chain of control actions $\bm{C}_{k-M}^k=(C_k,\bm{C}_{k-M}^{k-1})=(C_k,C_{k-1},\ldots,C_{k-M})$.

In the long-time regime, the joint probability of any sequence of control actions is invariant under $k$-translation, i.e., we can introduce
\begin{equation}
  \label{eq:Ps-def}
  P_{\sts}(\bm{\Gamma}^{M})\equiv P_{\sts}(\bm{C}^k_{k-M+1})=
  P_{\sts}(\bm{C}^{k'}_{k'-M+1}),
\end{equation}
where $\bm{\Gamma}^{M}\equiv \bm{C}^k_{k-M+1} = \bm{C}^{k'}_{k'-M+1}$ is a sequence of $M$ consecutive control actions. Note that our use of the subindex $\sts$ is consistent with the fact that $P_{\sts}(\bm{\Gamma}^{M})$ is a steady distribution, independent of $k$, i.e., independent of time. 
Therefore, $\Delta S_{c,s}^{(k)}$ is also independent of $k$: Using Eq.~\eqref{eq:deltaSc-k-indep-k} and Bayes theorem, we can write
\begin{align}
 \Delta S_{c,s}^{(k)} = -k_{B}\sum_{\bm{\Gamma}^{M+1}} P_{\sts}(\bm{\Gamma}^{M+1})\ln
  \frac{P_{\sts}(\bm{\Gamma}^{M+1})}{P_{\sts}(\bm{\Gamma}^{M})}, \; \forall M\ge M_0,
    \label{eq:deltaSc-k-indep-k-v2}
\end{align}
where, consistently,
$\bm{\Gamma}^{M+1} \equiv \bm{C}^k_{k-M} = \bm{C}^{k'}_{k'-M}$. 

Let us introduce now the dimensionless entropy $\Hseq(M)$ of a control sequence of 
length $M$ in the long-time regime,
\begin{equation}
  \label{eq:Scs-def}
  \Hseq(M)\equiv-\sum_{\bm{\Gamma}^{M}} P_{\sts}(\bm{\Gamma}^{M})\ln
  P_{\sts}(\bm{\Gamma}^{M}) .
\end{equation}
Equation \eqref{eq:deltaSc-k-indep-k-v2} can thus be rewritten as
\begin{equation}
    \Delta S_{c,s}^{(k)} = -k_{B}\dHs(M), \quad \forall M\ge M_0,
\end{equation}
where
\begin{align}
 \dHs(M)\equiv \Hseq(M+1)-\Hseq(M).
 \label{eq:Hp(M)}
\end{align}
The independence of $\Delta S_{c}^{(k)}$ with respect to $M$, for $M\ge M_0$, entails that $\dHs(M)=\dHs(M_0)$ and the entropy of a sequence of control actions $\Hseq$ is thus a linear function of the sequence length $M$ for large enough $M$,
\begin{equation}
\Hseq(M)=M \dHs(M_0)+H_0, \quad M\ge M_0,
\label{eq:Hs-linear}
\end{equation}
where $H_0$ is a constant.

The above property makes it possible to obtain the entropy change in
the long-time limit due to the updates of the control. We
take a long interval of length $\tau$, such that the control has
acted $N_{a}\gg 1$ times during it,
\begin{equation}
  \label{eq:tau-def}
  \tau=N_{a}\Delta t_{m}.
\end{equation}
The total entropy reduction of the particle due to the information gathered by the control is
\begin{align}
  \Delta S_{\info}&\equiv 
  \sum_{k=k_{r}+1}^{k_{r}+N_{a}}\Delta
  S_{pc}^{(k)}=
  -\sum_{k=k_{r}+1}^{k_{r}+N_{a}} \Delta S_{c,s}^{(k)} 
  \nonumber \\
  &= -N_{a}k_{B}\dHs(M_{0})\le 0.
  \label{deltaS-info}
\end{align}
The entropy reduction per measurement is thus
\begin{equation}
    \dSinfoPM = \frac{\Delta S_{\info}}{N_a} = - k_{B}\dHs(M_{0})\le 0, 
  \label{eq:dSinfoPM}
\end{equation}
and the entropy reduction rate is
\begin{equation}
    \label{eq:sigma-info}
    \sigma_{\info}\equiv \frac{\Delta S_{\info}}{\tau} 
    = \frac{\dSinfoPM}{\Delta t_m}
    \le 0.
\end{equation}
This is the rate at which the entropy of the particle is reduced in the long-time regime, due to the information gathered by the control. 

In the above, we have shown that if the memory of control values is restricted to the most recent $M_0$ values, then $\dHs(M)$ is constant for $M\ge M_0$. In other words, this memory loss is a sufficient condition for having a constant value of $\dHs(M)$. It can be shown, see Appendix~\ref{sec:app-dHs-const}, that this memory loss is also a necessary condition: if $\dHs(M)$ is constant for $M\ge M_0$, then the memory of control values is restricted to the most recent $M_0$ values.

\section{Thermodynamic balance}\label{sec:thermodyn-balance}

\subsection{First and second principles}\label{sec:1st-2nd-princ}

Now we bring to bear the first and second principles in the long-time
state. As in our discussion on the entropy reduction, we consider a
long time interval of duration $\tau$ such that the control has acted
a large number of times $N_{a}$, as given by Eq.~\eqref{eq:tau-def}.  All
increments (energy, entropy) in this section, as well as heat exchange
and work correspond to such a long time interval.

The first principle is the average energy balance for the particle, which reads
\begin{equation}
  \label{eq:first-principle}
  \cancelto{0}{\Delta\av{U}}=\av{W_{\inn}}+\av{W_{\out}}+\av{Q},
\end{equation}
in which $\av{W_{\inn}}$ stands for the average energy input by the control,
$\av{W_{\out}}$ is the average work done by the external force on the particle, and $\av{Q}$ is the average heat exchanged by the particle and the thermal bath. Note that both $\av{Q}$ and $\av{W}$ are positive (negative) when the average internal energy of the particle increases (decreases), and $\Delta\av{U}$ vanishes over the long interval of duration $\tau$.

The second principle deals with the entropy balance. There are two
contributions to the entropy change of the particle $\Delta S_{p}$:
first, we have the entropy change due to the updates of the control,
$\Delta S_{pc}$, and, second, the entropy change due to the
interaction with the thermal bath, $\Delta S_{pb}$. In the long-time regime,
\begin{equation}
  \label{eq:second-principle}
  \cancelto{0}{\Delta S_{p}}=\Delta S_{pc}+\Delta S_{pb}.
\end{equation}
The first term, $\Delta S_{pc}$, contributes at the control updates discrete times
$t_{k}=k\Delta t_{m}$---it is the one we have already analyzed in
Sec.~\ref{sec:entropy-def}. The second one, $\Delta S_{pb}$, contributes between
 control updates, i.e., in the time intervals
$J_{k}=(t_{k},t_{k+1})$. In these time intervals, the
particle exchanges heat with the heat bath, and then
\begin{equation}
  \label{eq:deltaS-pb-inequality}
  \Delta S_{pb}-\frac{\av{Q}}{T}\ge 0 \implies \av{Q}\le T\Delta S_{pb} = -T\Delta S_{pc}.
\end{equation}

\subsection{Efficiency of the ratchet}\label{sec:efficiency}

We recall that we are considering a long time interval of duration $\tau$, during which the control has acted a large number of times
$N_{a}=\tau/\Delta t_{m}$. In the long-time regime, the particle moves to the right with velocity
$v_{\sts}$, i.e., $d\av{x}/dt=v_{\sts}$, where $v_{\sts}$ is a constant, as long as $F_{\ext}<F_{\sto}$. Therefore, the particle performs work
against the external force,
\begin{equation}
    \av{W_{\out}}=-F_{\out} v_{\sts}\tau<0.
\end{equation}

Putting together the first and second principles, Eqs.~\eqref{eq:first-principle}--\eqref{eq:deltaS-pb-inequality}, we get
\begin{equation}
  \label{eq:1st-2nd-princ-together}
  -\av{W_{\out}}=\av{W_{\inn}}+\av{Q}\le \av{W_{\inn}}-T\Delta S_{pc}.
\end{equation}
Now we can substitute $\Delta S_{pc}$ with
$\Delta S_{\info}$ in Eq.~\eqref{deltaS-info},
\begin{equation}
  \label{eq:bound-W-out}
  -\av{W_{\out}}\le \av{W_{\inn}}-T\Delta S_{\info}.
\end{equation}
The rhs gives an upper bound for the work we can extract from the feedback controlled ratchet.

The efficiency of the ratchet may then be defined as
\begin{equation}
  \label{eq:eta-def}
  \eta\equiv \frac{-\av{W_{\out}}}{\av{W_{\inn}}-T\Delta S_{\info}},
\end{equation}
which is equivalent to Eq.~(16) in Ref.~\cite{cao_thermodynamics_2009}. Note that this efficiency is different from the \textit{open loop} efficiency
\begin{equation}
  \label{eq:eta0-def}
  \eta_{0}=\frac{-\av{W_{\out}}}{\av{W_{\inn}}}.
\end{equation}
In fact, we have $\eta<\eta_{0}$ because $\Delta S_{\info}<0$.  Note that $\eta_{0}$ may become larger than unity without any fundamental difficulty: it is essential to incorporate the entropy change due to the updates of the control $\Delta S_{\info}$ to have the actual efficiency of the ratchet---see also Sec.~\ref{sec:efficiency-num}.

We can write the efficiency in terms of power instead of work, by dividing both the numerator and the denominator of Eq.~\eqref{eq:eta-def} by the considered time span $\tau=N_{a}\Delta t_{m}$. Defining the output and input powers as
\begin{equation}
  \label{eq:powers}
  P_{\out}\equiv\frac{\av{W_{\out}}}{\tau}=-F_{\ext}v_{\sts}, \quad
  P_{\inn}\equiv\frac{\av{W_{\inn}}}{\tau}, 
\end{equation}
and recalling the definition of the entropy production rate, Eq.~\eqref{eq:sigma-info}, we can write
\begin{equation}
  \label{eq:eta-def-v2-powers}
  \eta\equiv \frac{|P_{\out}|}{P_{\inn}-T\sigma_{\info}}
\end{equation}
for the actual efficiency of the feedback controlled ratchet, whereas the open loop efficiency reads
\begin{equation}
  \label{eq:eta0-def-v2-powers}
  \eta_{0}=\frac{|P_{\out}|}{P_{\inn}}
\end{equation}

\section{NUMERICAL METHODS}\label{sec:numerical-methods}

\subsection{Simulation of the ratchet dynamics}\label{sec:simul-ratchet-dyn}

To simulate the feedback flashing ratchet dynamics, we discretize time by introducing a small time step $\delta t$ the Langevin equation, Eq.~\eqref{Eq:Langevin2}. Specifically, we employ the Euler algorithm,
\begin{equation}
x(t+\delta t) = x(t) - C_k\frac{V'(x(t))}{\gamma} \delta t - \frac{F_{\ext} }{\gamma} \delta t + z \sqrt{2 D \delta t} ,
\label{eq:Langevin-Euler}
\end{equation}
to integrate the Langevin equation between control updates, i.e., in each time interval $J_k=(t_k,t_{k+1})$. In Eq.~\eqref{eq:Langevin-Euler}, $D=k_B T/\gamma$ is the diffusion coefficient and $z$ is a Gaussian random number with  zero mean and unit variance, $\av{z}=0$, $\av{z^2}=1$. Units have been chosen as follows: The length unit is $L$ and the energy unit is $k_{B}T$, i.e., we take $L=1$ and $k_{B}T=1$, whereas the time unit is determined by taking $\gamma=1$---i.e., the time unit is given by $L^2/D=L^2\gamma /(k_B T)$.

At the discrete times $t_k=k\Delta t_m$, the state of the control is updated over each trajectory generated with Eq.~\eqref{eq:Langevin-Euler}. Before the control update, in the time interval $J_{k-1}=(t_{k-1},t_k)$, the state of the control was given by $C_{k-1}$, i.e. the particle felt the potential $C_{k-1}V(x)$. At $t=t_k$, the position of the Brownian particle is measured, and the potential is on (off) for the next time interval $J_k=(t_k,t_{k+1})$ if the force at this position $-V'(x(t_k))$ is positive (negative). In this way, we construct the sequence of control actions $\bmCk=(C_k,\ldots,C_1)$ for each trajectory.

The above procedure has been employed to build one trajectory of the dynamics of the particle, and the numerical results presented in this paper have been obtained by doing $N_{\simm}$ of such trajectories. For obtaining the time evolution of the  average physical quantities, we average over all the considered trajectories. In the long time regime, some quantities show time-translation invariance, and thus we perform the average over all the equivalent instances of one long trajectory. This additional time average can be done, for example, for the joint probability distribution of $M$ consecutive control actions, $P_{\sts}(\bm{\Gamma}^M)$, as detailed below.

\subsection{Numerical evaluation of the entropy reduction}\label{sec:eval-inf-entropy}

\begin{figure*}
  \centering
 \includegraphics[width=0.49\textwidth]{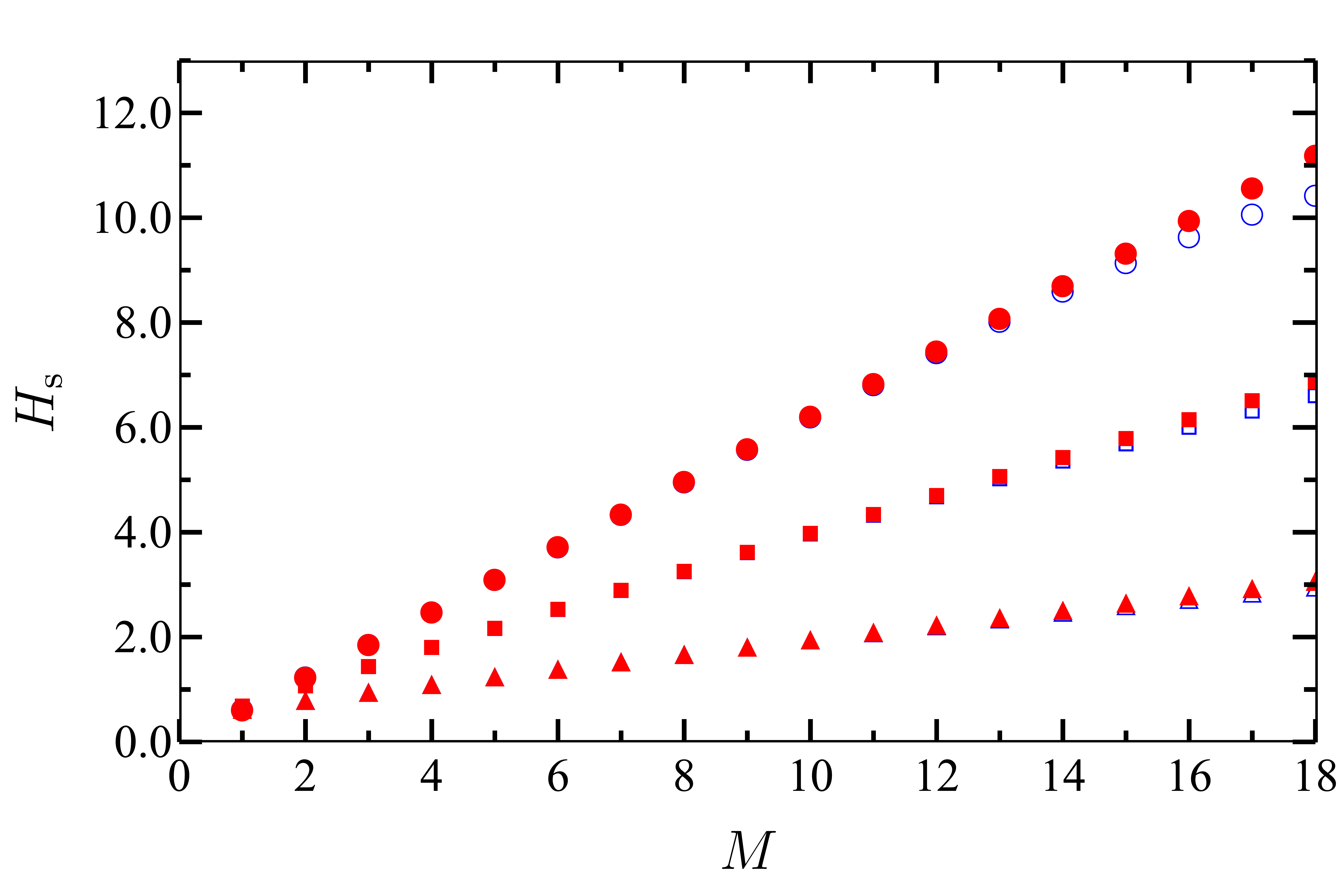}
 \includegraphics[width=0.49\textwidth]{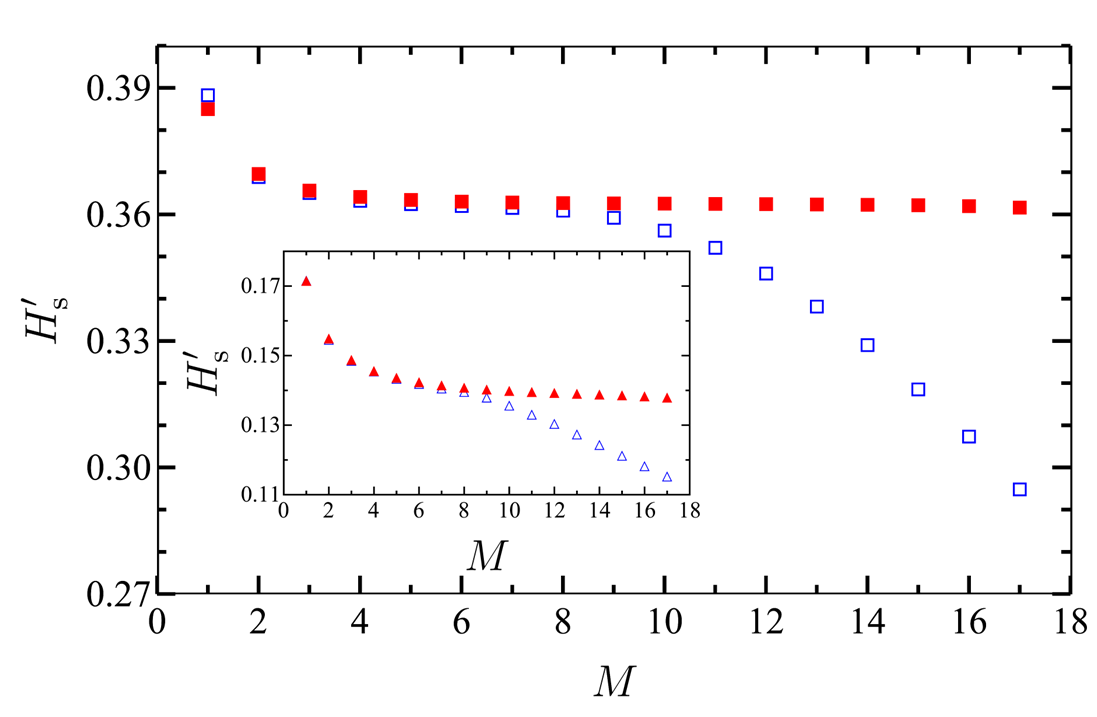}
  \caption{\label{fig:S-seq} 
  Convergence of the estimator for the entropy reduction per measurement $\dSinfoPM$.
  On the left panel, the entropy of sequence $\Hseq$ of control
    actions in the long-time regime  as a function of the length of the sequence $M$ is displayed, for data corresponding to one trajectory, i.e., $N_{\simm}=1$. Data for three different values of the time between control updates $\Delta t_m$ are presented, $\Delta t_m=0.02$ (circles), $\Delta t_m=0.002$ (squares), and $\Delta t_m=0.0002$ (triangles), with two values of the number of control actions $N_a$ for each case, $N_a=10^5$ (open blue symbols) and $N_a=10^8$ (filled red symbols). On the right panel, the corresponding slope $\dHs$ for $\Delta t_m=0.002$ is shown, together with the one corresponding to $\Delta t_m=0.0002$ (inset), using the same symbol and color code. The linear behavior of $\Hseq$, corresponding to a roughly constant value of $\dHs$, is neatly observed, beyond 
    $M\ge M_0\approx 4$ for $\Delta t_m=0.002$, and $M\ge M_0\approx 10$ for $\Delta t_m=0.0002$. System parameters are $a=1/3$ and $V_{0}=5$ (for the potential), and  $F_{\ext}=0$ (no external force). For these values of the parameters, the characteristic times of the system are: $0.5$ (diffusion time over $L$), $0.066$ (drag time in the $(0,a)$ interval), $0.133$ (drag time in the $(a,L)$ interval).
    (Units: $L=1$, $k_{B}T=1$, $\gamma=1$.)
    }
\end{figure*}
The discussion in Sec.~\ref{sec:long-time-entropy} allows us to extract the entropy reduction per measurement $\dSinfoPM$ from simulation data:
Eq.~\eqref{eq:dSinfoPM} gives $\dHs(M_{0})$ as an estimator for $-\dSinfoPM/k_B$.
In turn, Eqs.~\eqref{eq:Scs-def} and \eqref{eq:Hp(M)} provide us with a numerical procedure to evaluate $\dHs(M_{0})$, which requires estimating the probability 
$P_{\sts}(\bm{\Gamma}^{M})$ of sequences of control actions of length $M$.
Following our discussion in Sec.~\ref{sec:long-time-entropy},
$\dHs(M)$ is expected to become independent of $M$ for
$M\ge M_{0}$~\footnote{For a stationary stochastic process, $\dHs(M)$ is a monotonically decreasing function of $M$ with a well-defined limit for large $M$~\cite{cover_elements_2006}, this limit value equals $\dHs(M_0)$.}. If this expectation is fulfilled, the graph $\dHs(M)$
vs.~$M$ provides us with both $M_{0}$ and the value
$\dHs(M_{0})$. 

Despite the above, one has to take into account numerical limitations when sampling the probability distribution $P_{\sts}(\bm{\Gamma}^{M})$  over $N_{\simm}$ trajectories of a long duration $\tau=N_{a}\Delta t_{m}$. The number of possible control action sequences $\bm{\Gamma}^{M}$ is $2^{M}$, whereas the number of sequences of length $M$ for the consider sampling is $N_{\simm}(N_{a}-M+1)$. Therefore, it is necessary that $N_{\simm}(N_{a}-M+1)\gg 2^{M}$ or, taking into account that $N_{a}\gg M$, $N_{\tot}\equiv N_{\simm}N_{a}\gg 2^{M}$. For given values of $N_{\simm}$ and $N_{a}$, one expects to find a constant $\dHs(M)$ for $M>M_0$ until a certain point $M_{\max}(N_{\tot})$, for which the condition $N_{\tot}\gg 2^{M}$ is no longer fulfilled. From a practical point of view, one could estimate that $N_{\tot}/2^{M}\gtrsim 10^{2}$ to have a good sampling of the distribution $P_{\sts}(\bm{\Gamma}^{M})$. The accuracy of this qualitative picture can be checked by increasing the value of $N_{\tot}$  over which we sample the distributions $P_{\sts}(\bm{\Gamma}^{M})$ in the simulations, i.e., by increasing either the total length $\tau$ or the number of trajectories $N_{\simm}$. In this way, we should observe that $M_{\max}$ increases with $N_{\tot}$.

We have carried out the above described numerical procedure. We have chosen $a=1/3$ and $V_{0}=5$ for the potential parameters, in absence of external force, $F_{\ext}=0$. Figure~\ref{fig:S-seq} presents $\Hseq$  and $\dHs$ as a function of $M$ for three different values of the interval between control updates, $\Delta t_{m}=0.02$, $\Delta t_m=0.002$, and $\Delta t_m=0.0002$. Specifically, $\Hseq$ is plotted on the left panel, whereas the slope $\dHs(M)$, as defined in~Eq.~\eqref{eq:Hp(M)}, is plotted on the right panel.  A linear behavior of $\Hseq$ is neatly observed and, consistently, $\dHs$ is basically constant. There is a slight decrease of $\dHs$ for large $M$---almost not discernible in the graph of $\Hseq$, which is linked to the insufficient statistics for sampling $P_{\sts}(\bm{\Gamma}^{M})$ discussed above. We make this clear by considering two different values of $N_{a}$ (with $N_{\simm}=1$), which correspond to the two different sets of symbols on each panel: $N_{a}=10^{5}$ and $N_{a}=10^{8}$. The observed behavior is consistent with our previous discussion, $\dHs$ is not well sampled and thus decreases for $M>M_{\max}\approx \ln(0.01 N_a)/\ln 2$, $M_{\max}=10$ ($M_{\max}=20$) for $N_{a}=10^{5}$ ($N_a=10^8$).

The above plots of $\dHs(M)$ also allow us to identify $M_{0}$, the length of the history of control actions after which the memory is ``lost''. We recall that the constancy of $\dHs(M)$ for $M\ge M_{0}$ is linked to the conditional probability $P_{\sts}(C_{k}|C_{k-1},\ldots,C_{k-M+1})$ being independent of $M$ for $M\ge M_{0}$.  For $\Delta t_{m}=0.02$, we infer that $M_{0}\approx 2$ (not shown), for $\Delta t_m=0.002$, $M_0\approx 4$, and for $\Delta t_m=0.0002$, $M_0\approx 10$.  We have checked that a similar behavior is found for other values of the interval between control updates. The qualitative picture remains the same, but the value of $M_{0}$ above which $\dHs(M)$ becomes constant increases as $\Delta t_m$ decreases. This is sensible from a physical point of view: for a shorter interval between control updates, the correlation between adjacent control values increases. Therefore, it is necessary to consider longer chains to observe the decay of these correlations~\cite{cao_information_2009}\footnote{From a fundamental point of view, the fact that $M_{0}>1$ means that $\bm{\Gamma}^{M}$ is not a Markov process.}. 

Note that the same estimate for the entropy reduction can be obtained by considering
\begin{equation}
    \dSinfoPM = \frac{\Delta S_{\info}}{N_a} \simeq - k_{B}\frac{\Hseq(M_{0})}{M_0}\le 0, 
  \label{eq:dSinfoPM-seq}
\end{equation}
see for instance Refs.~\cite{cover_elements_2006, cao_information_2009}.
The convergence of this estimator is slower for the cases considered here, supporting the use of $-\dSinfoPM/k_B \simeq \dHs(M_0)$ throughout this work.

\subsection{Computational estimate of the efficiency}
 
Now we describe how we compute the efficiency of the feedback controlled ratchet in the simulations. The efficiency $\eta$ is given by Eq.~\eqref{eq:eta-def-v2-powers}, then we have to discuss how to estimate $P_{\inn}$ and $P_{\out}$ in the simulations---we have just discussed how to estimate $\sigma_{\info}$ in the previous section. 

On the one hand, we have that the output power is
$|P_{\out}|=F_{\ext} v_{\sts}$, as given by Eq.~\eqref{eq:powers}, 
with $v_{\sts}$ being the (constant) velocity of the particle in the long-time regime. This velocity is easily calculated in the simulations, by plotting $\av{x}$ as a function of time and getting the slope of this graph---after the initial transitory, i.e. for times $t>t_r=k_r\Delta t_m$.

On the other hand, the input power $P_{\inn}$ is provided by the change in the potential energy of the particle due to the on/off switching of the potential $V(x)$. Let us consider the $k$-th update of the control in a given trajectory, at time $t_k$ separating the time intervals $J_{k-1}$ and $J_k$. When the control turns on (off) the potential $V(x)$, the energy of the Brownian particle is increased (decreased) by an amount $V(x(t_k))$. Note that the on/off switching of the potential takes place only when $C_k\ne C_{k-1}$, if $C_k=C_{k-1}$ there is no energy change because the potential is either on (when $C_k=C_{k-1}=1$) or off (when $C_k=C_{k-1}=0$) in both time intervals $J_{k-1}$ and $J_k$. Therefore, for each considered trajectory one has 
\begin{equation}
    \Delta U^{(k)}= V(x(t_k)) \left( C_k - C_{k-1} \right).
\end{equation}
Note that $C_k - C_{k-1}=1$ ($C_k - C_{k-1}=-1$) when the potential is switched on (off).

As before, we consider long trajectories of duration $\tau$, as given by Eq.~\eqref{eq:tau-def}, once the system has reached the long time regime, i.e. $k>k_r$. Then, for each trajectory one has
\begin{equation}
    P_{\inn} = \frac{1}{N_a \Delta t_m}\sum_{k=k_r+1}^{k_r+N_a} \Delta U^{(k)},
\end{equation}
and this value is averaged over $N_{\simm}$ trajectories.

\subsection{Typical parameters employed in the simulations}

The time step $\delta t$ for integrating the Langevin equation has been chosen to be $\delta t=10^{-2}\Delta t_m$. For all the cases considered throughout the paper, this value of $\delta t$ is much smaller than all the characteristic time scales of the system, and therefore the continuous-time dynamics of the system is accurately reproduced in the numerics.

As we are interested in the long-time behavior of the system, we skip the transitory dynamics---i.e., we skip an initial time interval $t_r$, as described above. In this way, the system reaches the asymptotic long-time regime for $t>t_r=k_r\Delta t_m$, with $k_r\gg 1$: unless otherwise stated, we have taken $k_r=10^5$.
Afterwards, we simulate the system dynamics for a very long time $\tau$, as defined by Eq.~\eqref{eq:tau-def}. This time involves a very large number of control actions $N_a$, to gather sufficient statistics for the entropy reduction---as described above. 

We recall that our discussion in Sec.~\ref{sec:eval-inf-entropy} entails that the entropy reduction per measurement $\dSinfoPM$ is estimated from the entropy $\Hseq$ of sequences of control actions of length $M$ in the long-time regime. Specifically, it is given by the slope of $\Hseq$ vs. $M$ for a large enough value of $M$, $M\ge M_0$. For the range of $\Delta t_m$ considered in this paper, we have numerically checked that $M_0\le 10$. Thus we have fixed $M_0=10$ to compute the numerical estimate of the entropy reduction by information, i.e. in the simulations we take
\begin{equation}
\dSinfoPM = - k_B \dHs(10), \quad \sigma_{\info}=-k_B \frac{\dHs(10)}{\Delta t_m}.
\end{equation}

\section{RESULTS}\label{sec:results}

In this section, we first present a numerical verification of the existence of a long-time regime in the feedback flashing ratchet. Afterwards, we discuss the dependence of the entropy reduction per measurement $\dSinfoPM$ on the main parameters of the model: time between control updates $\Delta t_m$, potential asymmetry $a$, potential height $V_0$, and external force $F_{\ext}$. Also, we discuss the impact of the external force on the input and output powers $P_{\inn}$ and $P_{\out}$, showing that its quotient reveals the need of incorporating the contribution of the entropy reduction rate $\sigma_{\info}=\dSinfoPM/\Delta t_m$ to the efficiency definition, as given by Eq.~\eqref{eq:eta-def-v2-powers}.

\subsection{Long-time regime}\label{sec:results-long-time}

Here we show numerically the existence of the long-time regime, in which the Brownian particles move in average to the right with a constant velocity, whereas other averaged physical quantities reach a periodic behavior, with period equal to the time between control updates $\Delta t_m$.

\begin{figure}
    \centering
    \includegraphics[width=1\linewidth]{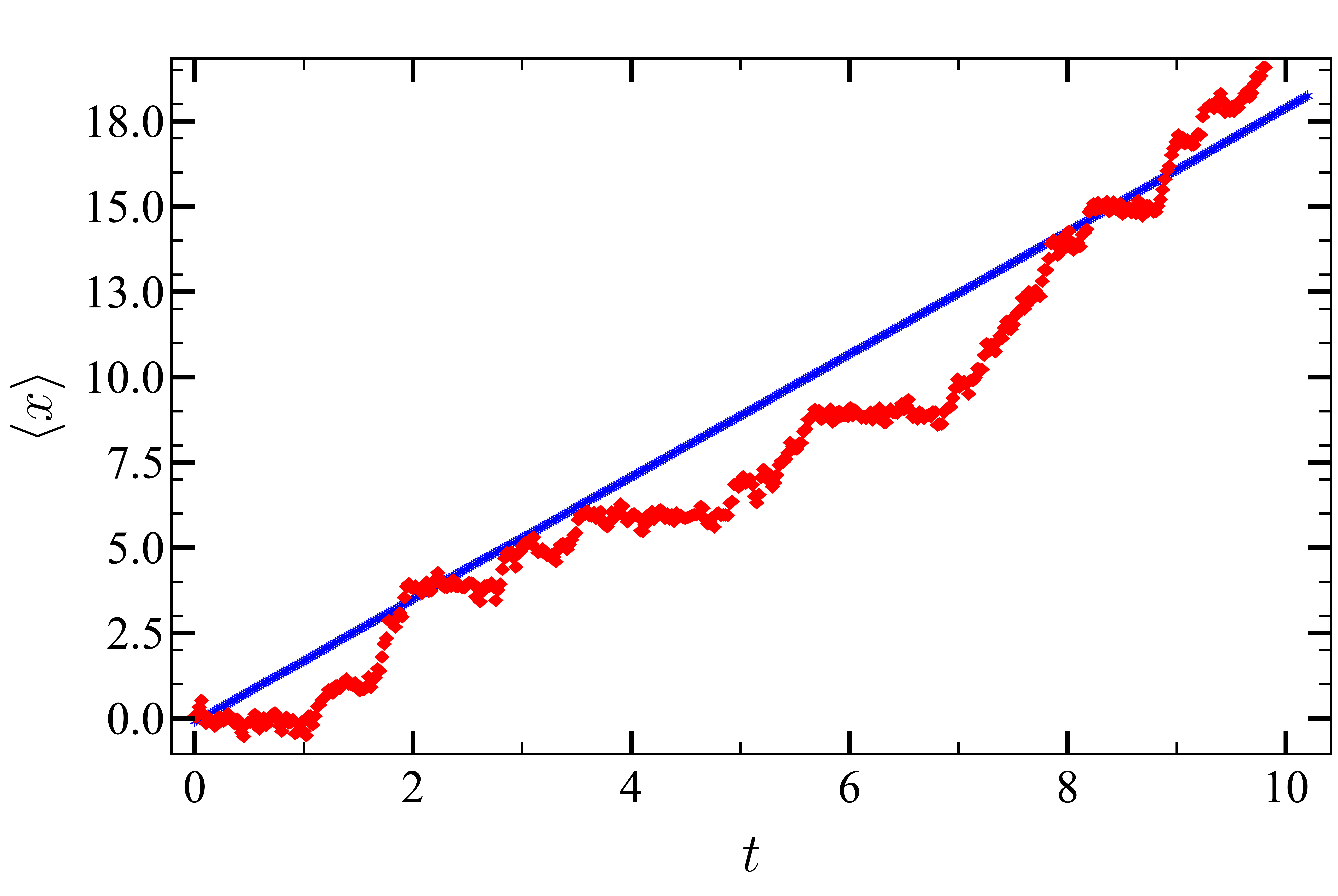}
    \caption{ Time evolution of the particle's position and its average value. Specifically, we plot $\langle x (t)\rangle $ (blue solid line), averaged  over $N_{\text{sim}}=10^5$ trajectories, and $x(t)$ (red squares) for a single trajectory. Due to the action of the feedback controller, the particle flows to the right. The system parameters are $\Delta t_{\text{med}}=0.02, F_{\text{ext}}=0, V_0=5, a=1/3$.
    (Units: $L=1$, $k_{B}T=1$, $\gamma=1$.)
    }  
    \label{fig:x_t}  
\end{figure}

Figure~\ref{fig:x_t} illustrates the directed flow of the system. We present the average value, over $N_{\simm}=10^5$ trajectories, of the position $\av{x}$ as a function of time, together with one of the trajectories, for the typical set of parameters we consider in this numerical section: $\Delta t_m=0.02$, $F_{\ext}=0$, $V_0=5$, and $a=1/3$. In the individual trajectory, it is clearly observed that the motion of the particle comprises two main regimes: Basically, there are time intervals in which the particle (i) is localized, fluctuating around a fixed position, and (ii) travels at constant speed to the right. Regime (i) corresponds to the thermal fluctuations of the particle, typically centered around a point to the right of the minima (due to the action of the control, which switches off the potential if the particle is found in this region) of $V(x)$. This random motion continues until one fluctuation is large enough so as to transfer the particle to the next periodic repetition of the potential (typically to the right, which is closer), and thus regime (ii) emerges, with the particle sliding down to the next minimum. For the set of parameters employed, the range of times shown corresponds to $500$ updates of the control.

\begin{figure}
    \centering
    \includegraphics[width=1\linewidth]{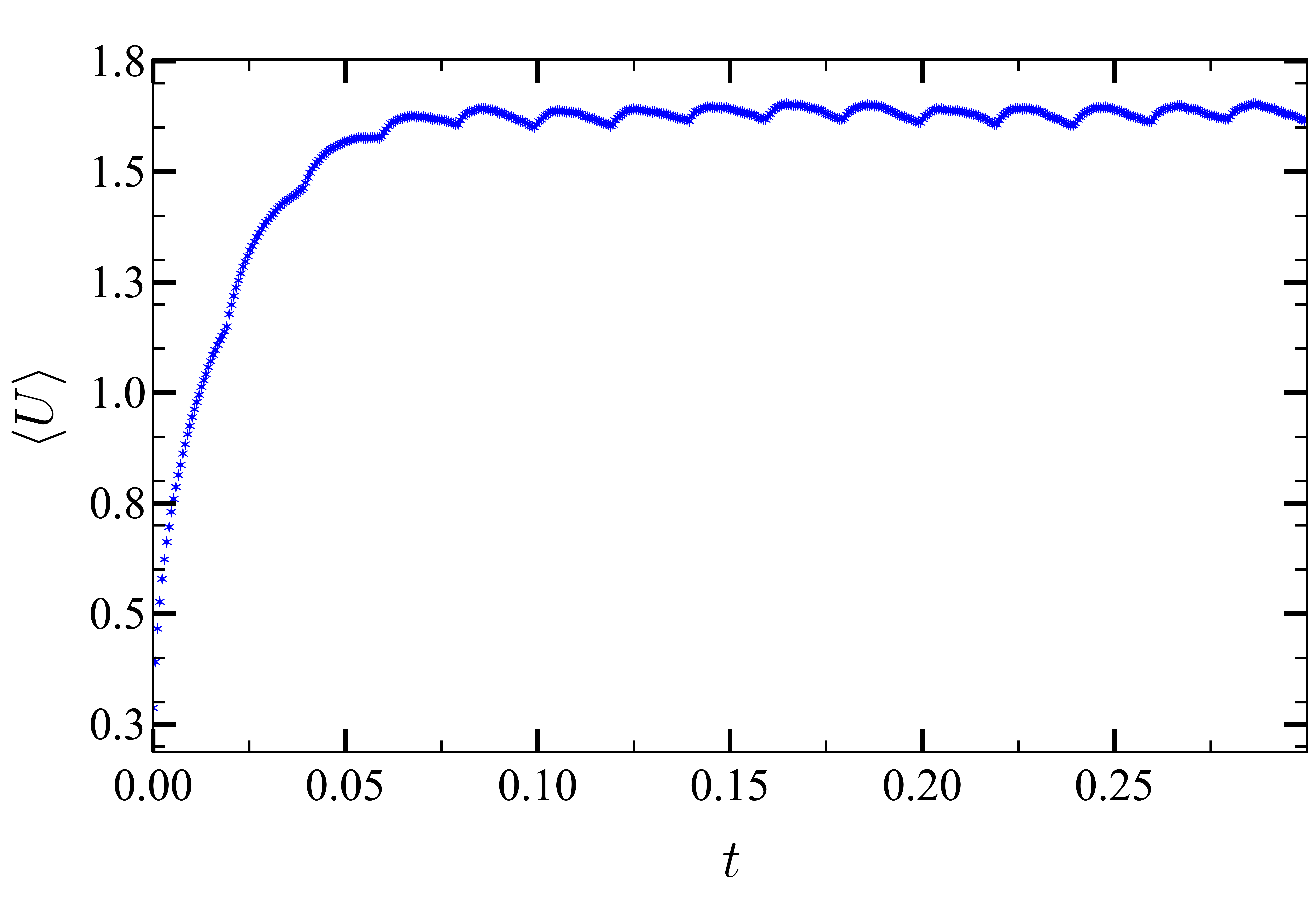}
    \caption{Time evolution of the particle's potential energy $\langle U \rangle $. The average has been performed over $N_{\text{sim}}=10^5$ trajectories. The asymptotic periodic behavior of $\av{U}$ is clearly observed. System parameters and units are the same as in Fig.~\ref{fig:x_t}.
    }  
    \label{fig_V_t}
\end{figure}

The time evolution of the average potential energy $\av{U}$ is shown in Figure~\ref{fig_V_t}, for the same set of parameters in Fig.~\ref{fig:x_t}. After a short transient, roughly corresponding to $5$--$6$ updates of the control, $\av{U}$ reaches a periodic regime. It is in this long-time regime that we apply the thermodynamic balance in Sec.~\ref{sec:1st-2nd-princ}. As compared with the previous figure, the considered time interval is shorter, corresponding to only $15$ updates of the control, to allow discerning the initial transient.  

\subsection{Entropy reduction}

The entropy of a sequence of control actions $\Hseq$ is a linear function of the sequence length $M$ for large enough $M\ge M_0$, as given by Eq.~\eqref{eq:Hs-linear}. Since the entropy reduction by information is proportional to the slope $\dHs(M_0)$, the entropy reduction would be larger (smaller) for those situations in which the entropy of the sequence $\Hseq$ increases (decreases). There are two extreme situations, which can be used to understand the behavior of the entropy reduction as a function of the system parameters. First, $\Hseq(M)$ is maximized when the probability $P_{\sts}(\bm{\Gamma}^M)$ is flat over all sequences $\bm{\Gamma}^M$, i.e., $P_{\sts}(\bm{\Gamma}^M)=2^{-M}$, whereas $\Hseq(M)$ is minimized when $P_{\sts}(\bm{\Gamma}^M)$ is peaked at a single sequence, i.e., when the evolution of the control becomes deterministic.

The above discussion entails that $\dSinfoPM$ increases when the probability of the values of the control, $C_k=1$ (on) and $C_k=0$ (off), become more symmetric as a consequence of changing the system parameters. In other words, $\dSinfoPM$ is expected to increase when the probability of finding the particle to the right and to the left of the potential minima is symmetrized~\footnote{The absolute maximum $k_B \ln 2$ may not be reachable for all sets of parameters, because these probabilities may not reach $1/2$.}. On the contrary, $\dSinfoPM$ decreases when the probability of the values of the control become less symmetric, vanishing in the limit when  the probability of finding the particle either to the left or to the right of the minima tends to unity---corresponding to control sequences $\bm{\Gamma}^M=(1 1 1 \ldots)$ and $\bm{\Gamma}^M=(0 0 0 \ldots)$, respectively. 

It must be highlighted that $\dSinfoPM$ is also expected to decrease when both values of the control are possible but there are very few possible sequences, due to strong correlations between consecutive control values. For example, if all particles were moving to the right at constant speed, without any thermal fluctuations, the value of the control would be perfectly enslaved to the particle position and only one control sequence (and its cyclic permutations) would be observed, which results in a vanishing $\dSinfoPM$.

\subsubsection{As a function of control update time $\Delta t_{m}$} \label{sec:S-info-func-tm}

\begin{figure}
    \centering
    \includegraphics[width=1\linewidth]{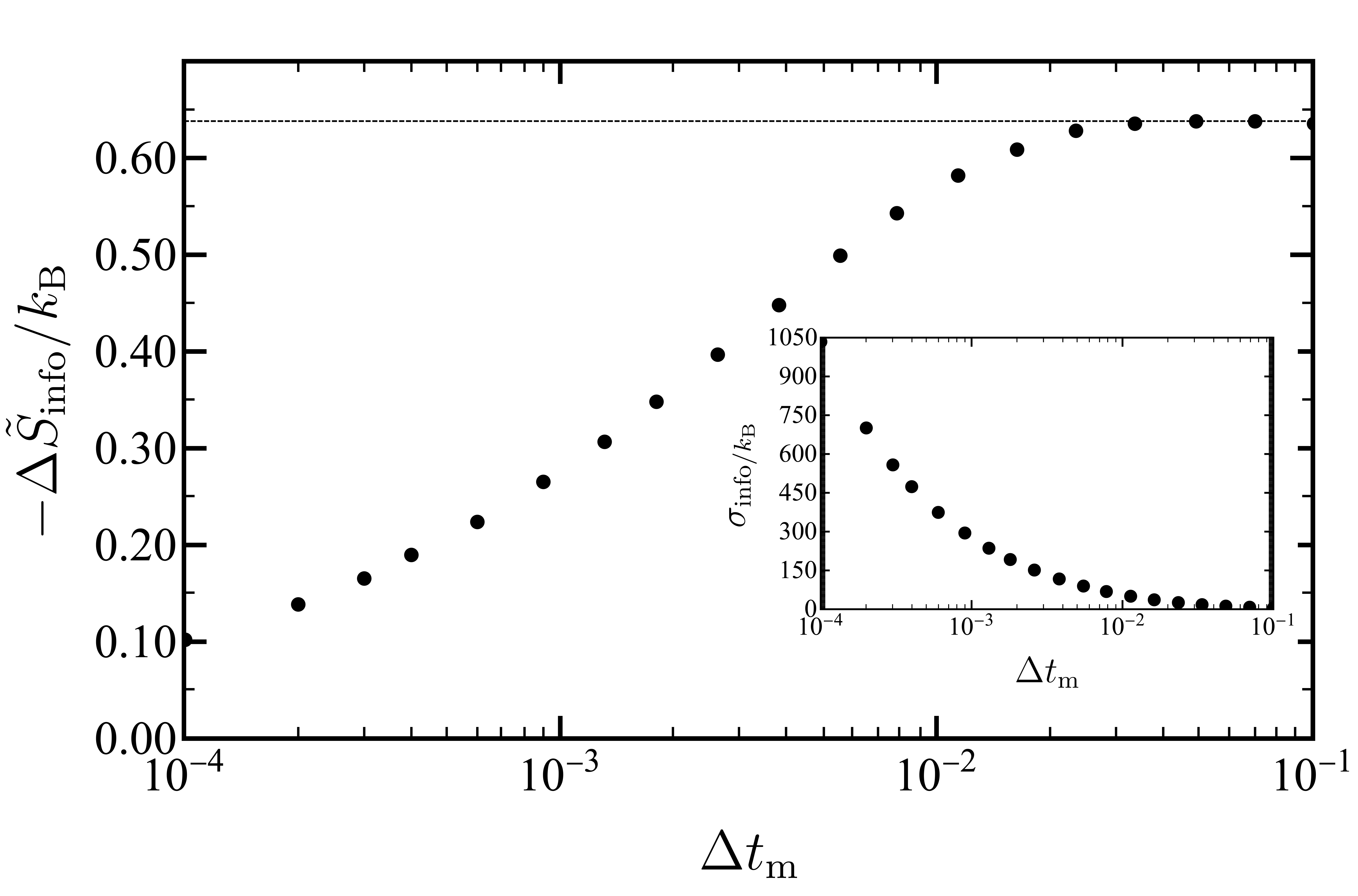}
    \caption{Entropy reduction  {per measurement} $\dSinfoPM$ 
    as a function of the time between control updates $\Delta t_{m}$.
     The inset shows the entropy reduction rate 
    $\sigma_{\info}$. System parameters and units are the same as in Fig.~\ref{fig:S-seq},  $V_0=5$, $a=1/3$, and $F_{\ext}=0$. $\dSinfoPM$ monotonically increases, reaching the asymptotic value $\Delta S_{\asy}$ in Eq.~\eqref{eq:DeltaS-asy} (dashed line) when $\Delta t_m$ becomes of the order of the longest characteristic time---see Fig.~\ref{fig:S-seq} for the values of the characteristic times.
    }
    \label{fig:entropia_tmed}
\end{figure}
The entropy reduction increases with the time between control updates. This is neatly shown in Fig.~\ref{fig:entropia_tmed}, where it is also observed that $\dSinfoPM$ saturates for long enough $\Delta t_m$. More specifically, $\dSinfoPM$ becomes constant when $\Delta t_m$ is of the order of the longest characteristic time for the particle.

For $\Delta t_{m} \to 0$, the particle is effectively moving under the action of the effective potential shown in Fig.~\ref{fig:potencial-efectivo}. The value of the control variable $C_k$ is thus basically enslaved to the previous particle position, since the latter has practically no time to evolve between control updates. This reduces the variety of possible sequences, giving rise to small values of $\dHs(M_0)$, i.e., small values of $\dSinfoPM$.

As $\Delta t_m$ increases, the control variable $C_k$ is no longer enslaved to the previous particle position and thus the number of possible sequences increases. When $\Delta t_m$ becomes longer than the characteristic relaxation times of the evolution under the on and off potentials, the particle position distribution reaches the corresponding equilibrium---in the absence of external force---before the next update of the control. This implies that the probabilities of the measurement output become independent of $\Delta t_m$. Consistently, $\dSinfoPM$ reaches a plateau, getting its maximum possible value---as depicted in Fig.~\ref{fig:entropia_tmed}.

The behavior found for $\Delta t_m\to\infty$ can be understood as follows. In the absence of external force, the system would reach the equilibrium state corresponding to the potential $V(x)$ in each time interval $J_k$ and thus consecutive values of the control, i.e., $C_{k-1}$ and $C_k$, become independent. This means that $M_0=1$ and $\dSinfoPM$ is directly linked to the stationary distribution $P_{\sts}(C)$ of control values. Furthermore, it can be easily checked that the probability of finding the particle at each side of the minimum---in the $x$ modulo $L$ problem, i.e. with periodic boundary conditions---is proportional to the respective lengths. 
Then, the corresponding probabilities of the values of the control are $P_{\sts}(C=0)=a$ and $P_{\sts}(C=1)=1-a$. Consistently, the entropy reduction $\dSinfoPM/k_B$ is expected to approach the asymptotic value
\begin{equation}\label{eq:DeltaS-asy}
   \frac{\Delta S_{\asy}}{k_B} = a \ln a + (1-a)\ln (1-a). 
\end{equation} 
The correctness of this intuitive picture is confirmed by the numerical value obtained in our simulations: In Fig.~\ref{fig:entropia_tmed} we have that $a=1/3$ 
and $\dSinfoPM/k_B$ agreeingly approaches $-0.637$.

Instead, the entropy reduction rate per unit of time 
$\sigma_{\info}$, as defined in Eq.~\eqref{eq:sigma-info}, increases as $\Delta t_m$ decreases, as observed in the inset of Fig.~\ref{fig:entropia_tmed}. This is sensible from a physical point of view: let us recall the definition of efficiency, Eq.~\eqref{eq:eta-def-v2-powers}, into which the entropy reduction rate enters. In the limit as $\Delta t_m \to\infty$, the feedback control plays no role and consistently $\sigma_{\info}\to 0$, and the actual efficiency $\eta$ and the open-loop efficiency $\eta_0$ coincide. In the limit as $\Delta t_m\to 0$, the role of the feedback controller is most important, so $\sigma_{\info}$, and also the difference between $\eta_0$ and $\eta$, is maximized---see Sec.~\ref{sec:efficiency} for a more detailed discussion.

\subsubsection{As a function of the potential asymmetry $a$}

\begin{figure}
    \centering
    \includegraphics[width=1\linewidth]{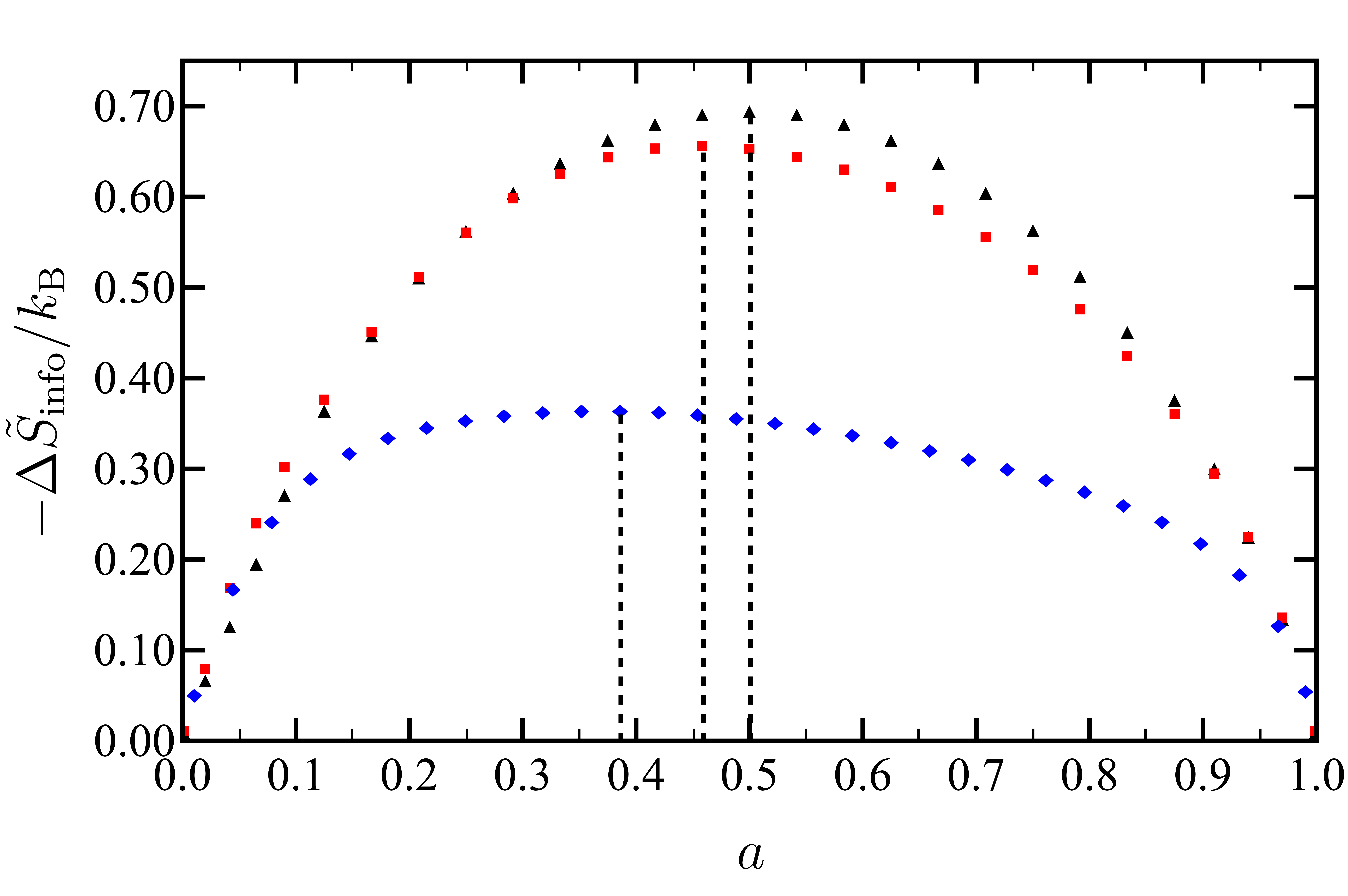}
    \caption{Entropy reduction  per measurement $\dSinfoPM$  as a function of the potential asymmetry $a$.  The entropy reduction vanishes for both $a\to 0$ and $a\to 1$, when the off and on control actions dominate, respectively. Three sets of points, corresponding to different values of the time between control updates, $\Delta t_m=0.2$ (black triangles), $\Delta t_m=0.02$ (red squares), and $\Delta t_m=0.002$ (blue diamonds), are shown. Vertical dashed lines indicate maxima position.
    Other system parameters are $V_0=5$ and $F_{ \text{\ext}}=0$. 
    }
    \label{fig:entropia_a}
\end{figure}
The entropy reduction shows a nonmonotonic behavior as a function of the potential asymmetry. This is clearly observed in Fig.~\ref{fig:entropia_a}, where we show $\dSinfoPM$ as a function of $a$ for the case $V_0=5$, $F_{\ext}=0$, and three different values of $\Delta t_m$, namely $0.2$, $0.02$, and $0.002$.

In all cases, the minima of $\Hseq$ are located on the extreme cases of totally asymmetric potentials, $a=0$ or $a=1$, where the force is positive or negative everywhere. This clearly leads to completely deterministic control sequences,  with all values $C_k=1$ (for positive force, $a=0$) or $C_k=0$ (for negative force, $a=1$): $\Hseq$, and thus $\dSinfoPM$, vanishes for both $a\to 0$ and $a\to 1$. Since $-\dSinfoPM$ is nonnegative, it must attain a maximum for an intermediate value of $a$. 
In Fig.~\ref{fig:entropia_a}, for the values of the parameters considered, the maximum is located at $a=1/2$ for the longest time between control updates, $\Delta t_m=0.2$. As the value of $\Delta t_m$ decreases, the maximum of $-\dSinfoPM$ decreases in height and moves to the left. In order to understand the observed behavior, it is instructive to consider the situation for a very long and very short time between control updates, i.e., $\Delta t_m\to\infty$ and $\Delta t_m\to 0^+$.

In the limit as $\Delta t_m\to\infty$, as already stated in Sec.~\ref{sec:S-info-func-tm}, the probability of finding the particle at both sides of the minimum is simply proportional to their respective lengths. Therefore, these probabilities are equal for the case $a=1/2$, and at this point $-\dSinfoPM$ attains its maximum. Also, this maximum equals $\Delta S_{\asy}$ in Eq.~\eqref{eq:DeltaS-asy}, since consecutive values of the control are not correlated. 

In the limit as $\Delta t_m\to 0^+$, the particle is moving under the effective potential depicted in the bottom panel of Fig.~\ref{fig:potencial-efectivo}. For $V_0\gg k_B T$, the particle freely diffuses in the interval $(0,a)$ whereas it moves ballistically with velocity $V_0/(\gamma L (1-a))$ in the interval $(a,L)$. The probability of finding the particle in both intervals can be estimated to be proportional to the time spent in them, i.e., $p_L \propto a^2/(2D)$ ($p_R\propto \gamma L^2 (1-a)^2/V_0$) for the probability of finding the particle with $0\leq x/L \leq a$ ($a\leq x/L \leq 1$). Both probabilities would be equal when $a^2/(1-a)^2=2k_BT/V_0$, which is smaller than $a=1/2$ in the considered regime  and tends to zero for $V_0/k_B T\to\infty$~\footnote{In the opposite limit, for $V_0\ll k_B T$, the effect of the confinement is negligible and the particle freely diffuses over the whole real line. Then, the maximum of $-\dSinfoPM$ would be located very close to $a=1/2$ for all $\Delta t_m$.}. Then, on an intuitive basis, we expect the maximum of $-\dSinfoPM$ to move to the left as $\Delta t_m$ decreases, in agreement with Fig.~\ref{fig:entropia_a}. Note that the height of the maximum decreases with $\Delta t_m$, which is consistent with the behavior of $-\dSinfoPM$ as a function of $\Delta t_m$ presented in Fig.~\ref{fig:entropia_tmed} and the behavior of $\dHs(M)$ in Fig.~\ref{fig:S-seq}; as $\Delta t_m$ decreases, consecutive values of the control become more and more correlated and the entropy reduction becomes smaller.

\subsubsection{As a function of the potential height $V_0$}

\begin{figure}
    \centering
    \includegraphics[width=1\linewidth]{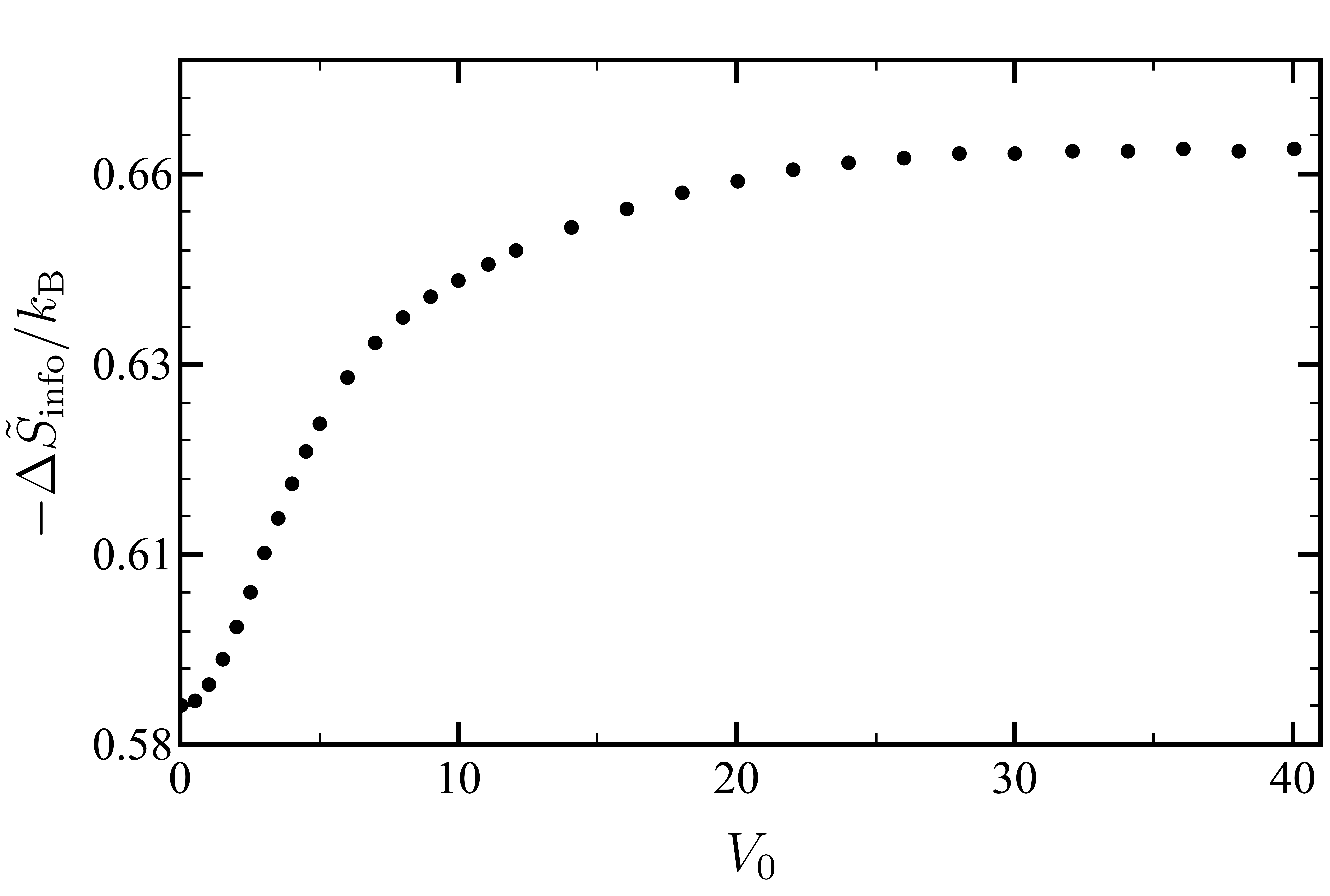}
    \caption{Entropy reduction per measurement $\dSinfoPM$ as a function of the potential height $V_0$. The entropy reduction increases, saturating for potential height values $V_0 \gtrsim 30$. System parameters are $ \Delta t_{ \text{m}}=0.02$, $a=1/3$, and $F_{ \text{\ext}}=0$.
     }
    \label{fig:entropia_V}
\end{figure}
The entropy reduction increases as a function of $V_0$ and saturates for large potential heights. This behavior is illustrated in Fig.~\ref{fig:entropia_V}. As $V_0$ is increased, the particle is effectively more and more confined to a narrow region around the potential minima, with small fluctuations due to the thermal noise. Therefore, the larger $V_0$ is, the more alike the probabilities of the two possible values of the control become. This (imperfect) symmetrization of the control values imply an increase of $\Hseq$ and thus of $-\dSinfoPM$. 

In the limit as $\Delta t_m\to\infty$, we recall that the particle reaches a steady state between control updates where $P_{\sts}(C=0)=a$, $P_{\sts}(C=1)=1-a$, independent of $V_0$, in the absence of external force. Therefore, for very long $\Delta t_m$, $\dSinfoPM$ would be basically independent of $V_0$. As $\Delta t_m$ decreases, the range of values swept by $\dSinfoPM$ as a function of $V_0$ increases, being maximum for $\Delta t_m\to 0^+$, where the particle effectively moves under the action of the effective potential depicted in Fig.~\ref{fig:potencial-efectivo}.

\subsubsection{As a function of the external force $F_{\ext}$}  

\begin{figure}
    \centering
    \includegraphics[width=1\linewidth]{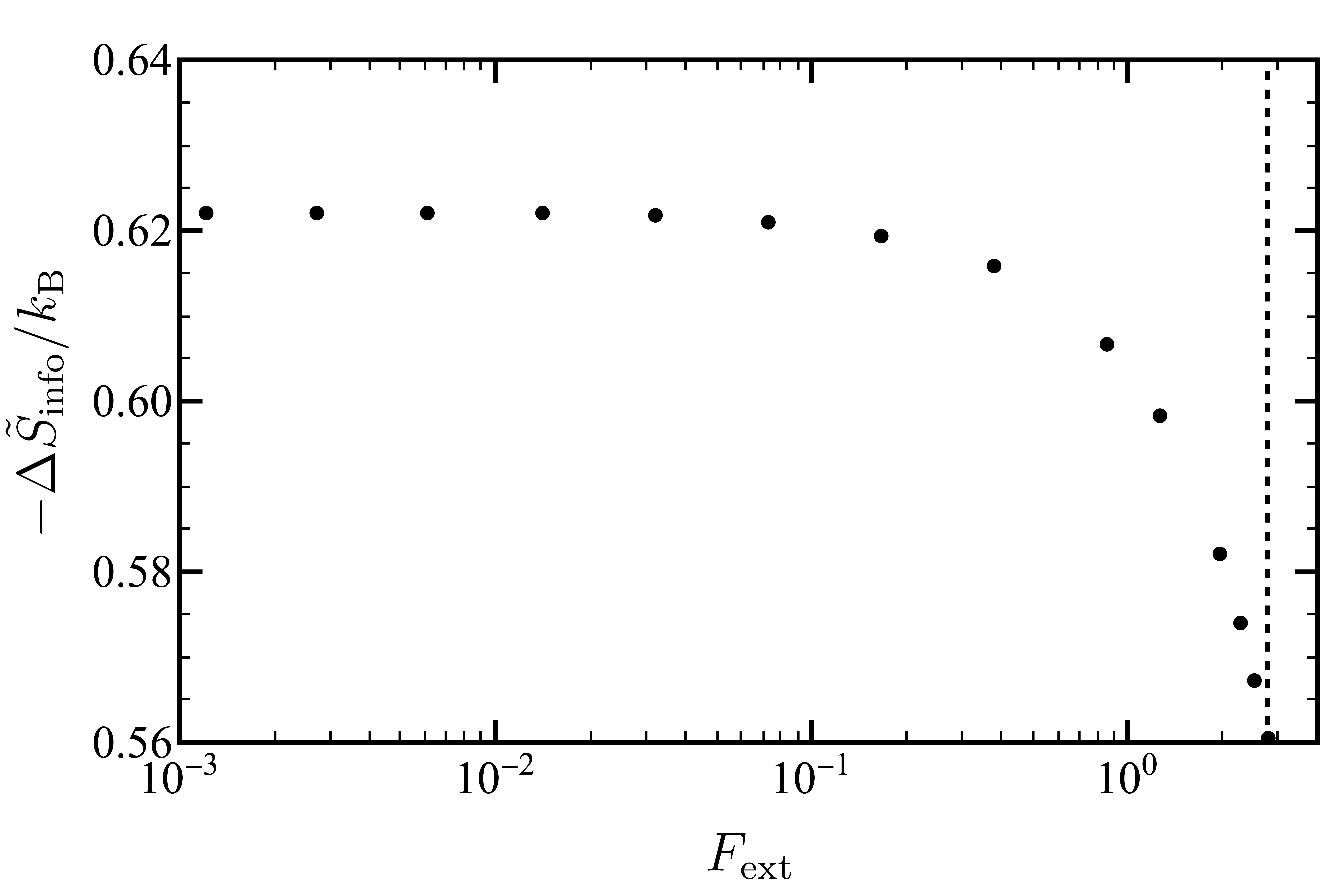}
    \caption{Entropy reduction  {per measurement $\dSinfoPM$ } as a function of the external force $F_{\text{\ext}}$. System parameters are $ \Delta t_{m}=0.02$, $a=1/3$, and $V_0=5$. Increasing the external force decrease the entropy reduction, particularly for values $F_{\ext} > 1 $---the force unit is given by the quotient of energy and length units, i.e. the force unit is  $k_B T / L$.
   }
    \label{fig:entropia_F}
\end{figure}
Figures~\ref{fig:entropia_tmed}-\ref{fig:entropia_V} for the entropy reduction correspond to the particle moving in the absence of an external force, $F_{\ext}=0$. Now we consider the influence of the external force. In Fig.~\ref{fig:entropia_F}, it is shown that the entropy reduction per measurement  $-\dSinfoPM$ decreases as the external force tends to $F_{\sto}$---at which the flux vanishes. We recall that the external force acts constantly on the particle, even if the potential is switched off, pushing the particle against the direction of flow. 

Increasing the external force increases (decreases) the confinement of the particles to the right (left) of the minima of the potential, where the force $-V'$ opposes (favors) the average flow to the right. Therefore, the probability of the potential being switched off (on)  decreases (increases).  This loss of symmetry of the on and off values of the control entails that $\Hseq$ and thus $-\dSinfoPM$ decreases with $F_{\ext}$.

\subsection{Efficiency}\label{sec:efficiency-num}

\begin{figure}
    \centering
    \includegraphics[width=1\linewidth]{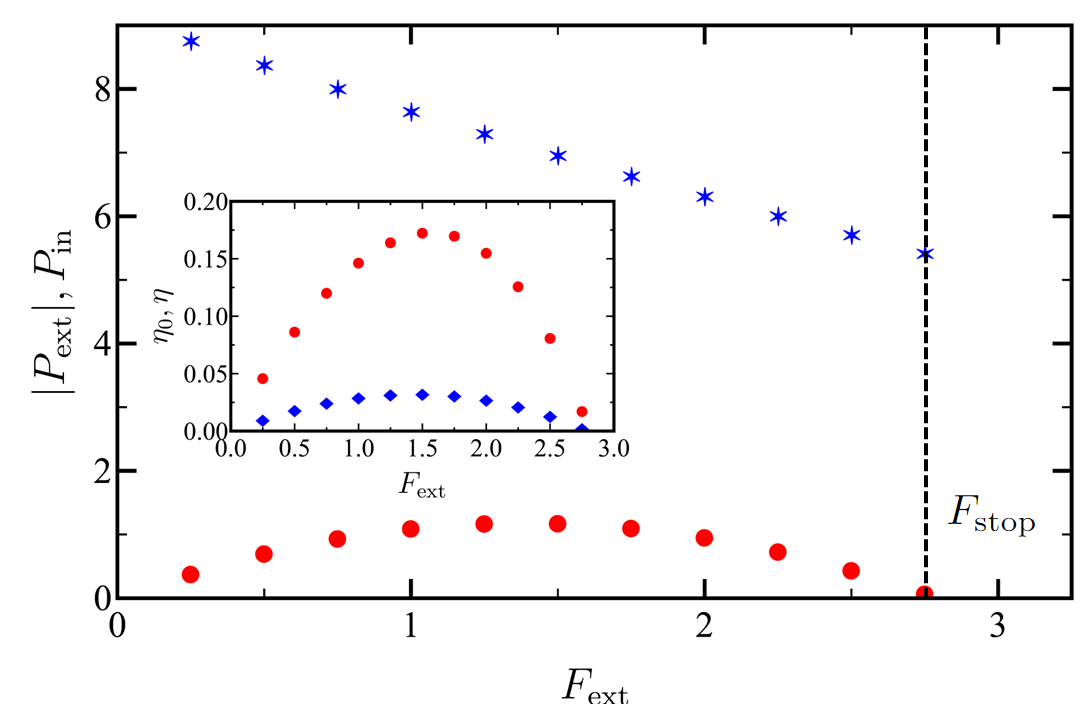}
    \caption{
    Input and output powers, $P_{\inn}$ and $P_{\out}$, as a function of the external force $F_{\text{\ext}}$. Specifically, we plot $P_{\inn}$ (blue stars) and $|P_{\out}|$ (red circles, recall that $P_{\out}<0$), averaged over one trajectory corresponding to a long time with $N_a=10^8$ control updates. In the inset,  the open-loop efficiency $\eta_0$ (red circles) and the actual efficiency $\eta$ (blue diamonds) are shown. System parameters are identical to those in Fig.~\ref{fig:entropia_F}, for which $F_{\sto}\approx 2.75$.
    }  
    \label{fig:termo_f2}    
\end{figure}
\begin{figure}
    \centering
     \includegraphics[width=1\linewidth]{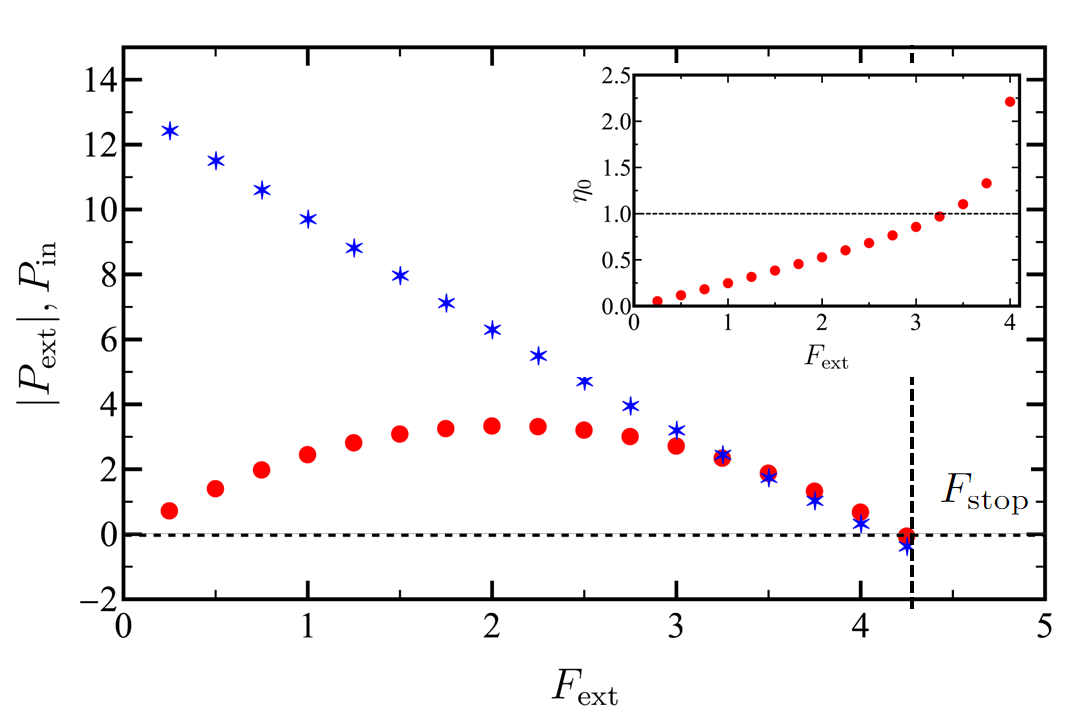}
    \caption{
    Same as in Fig.~\ref{fig:termo_f2}, but for a smaller time between control updates. Specifically, we are showing data corresponding to $\Delta t_m=0.002$, for which $F_{\sto}\approx 4.25$. For $F_{\ext}\gtrsim 3.5$, $P_{\inn}<|P_{\out}|$ and thus $\eta_0>1$, as observed in the inset.
    }  
    \label{fig:termo_f}
\end{figure}
Here, we present the numerical results for the efficiency, as well as for the output power $P_{\out}$---developed by the system against the external force---and the input power $P_{\inn}$---the rate at which energy is injected into the system due to the control updates. These two quantities are necessary  for the calculation of the efficiency, both the open loop efficiency $\eta_0$, Eq.~\eqref{eq:eta0-def-v2-powers}, and the actual efficiency $\eta$, Eq.~\eqref{eq:eta-def-v2-powers}, which in addition needs the entropy reduction. 

Figure~\ref{fig:termo_f2} shows the powers $P_{\inn}$ and $P_{\out}$ as a function of the external force $F_{\ext}$, for $F_{\ext}<F_{\sto}$. For the values of the parameters employed in the figure,  $F_{\sto} \approx 2.75$,  at this value of the external force the average flow is canceled---and thus $v_{\sts}$ and $P_{\out}$ vanish. Note that the output power $P_{\out}$ also vanishes for $F_{\ext}\to 0$, so it displays a maximum at a intermediate value of the external force. On the other hand, the input power $P_{\inn}$ by the control monotonically decreases as a function of $F_{\ext}$. In the inset, we show both the open-loop efficiency $\eta_0$ and the actual efficiency $\eta$. On an intuitive basis, $\eta_0$ seems to be well-behaved: since $|P_{\out}|<P_{\inn}$ over the whole range of forces, $\eta_0$ remains smaller than unity. Still, the actual efficiency $\eta$ is quite smaller than $\eta_0$, due to the contribution of the entropy reduction. The fundamental role played by $\dSinfoPM$ when computing the efficiency of the ratchet is further clarified by Figs.~\ref{fig:termo_f} and~\ref{fig:eta_f} below. 

Figure~\ref{fig:termo_f} is similar to Fig.~\ref{fig:termo_f2}, but for a smaller--by a factor of $10$---value of the time between control updates $\Delta t_m$. The qualitative behavior of $P_{\inn}$ and $P_{\out}$ as a function of $F_{\ext}$ remains basically the same, but an important difference arises: there appears a range of forces, $F_{\ext}\in[3.5, F_{\sto} \approx 4.25]$, for which the output power against the external force becomes larger (in absolute value) than the input power by the control. This means that the open loop efficiency (shown in the inset) becomes larger than unity---making crystal clear that this definition of the efficiency is not correct for feedback controlled systems.

Figure~\ref{fig:eta_f} presents the actual efficiency of the feedback controlled ratchet, as given by Eq.~\eqref{eq:eta-def-v2-powers}, also as a function of $F_{\ext}$, for the same value of $\Delta t_m$ considered in Fig.~\ref{fig:termo_f}. As opposed to the open-loop efficiency, its value remains bounded between $0$ and $1$---the physical reason is the incorporation of the entropy reduction by information to the energy balance of the system, as discussed in Sec.~\ref{sec:efficiency}.  Qualitatively, the behavior of the efficiency is very similar to that of $P_{\out}$, vanishing for both $F_{\ext}\to 0$ and $F_{\ext}\to F_{\sto}$.
\begin{figure}
    \centering
    \includegraphics[width=1\linewidth]{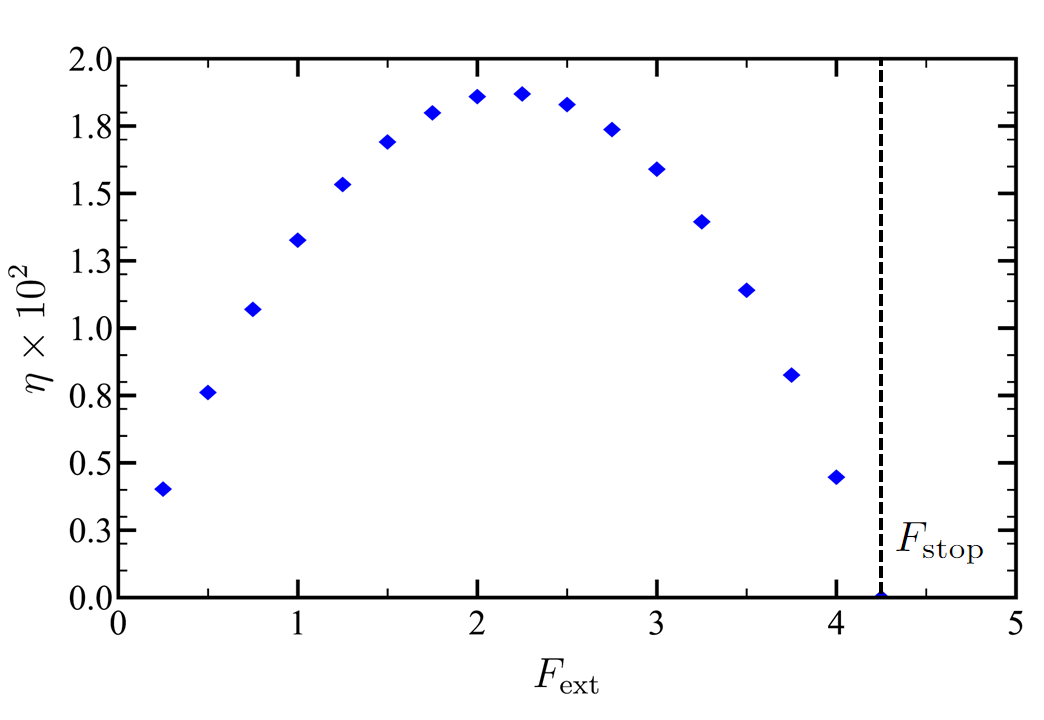}
    \caption{Actual efficiency $\eta$ of the feedback controlled ratchet as a function of the external force $F_{\ext}$. System parameters are the same as in the previous figure. This efficiency properly accounts for the reduction of the particle entropy due to the information gathered by the control, and thus it is consistently smaller than unity. 
    }
    \label{fig:eta_f}
\end{figure}

 \section{CONCLUSIONS}\label{sec:conclusions}
 
In this work, we have investigated the entropy reduction by information of a Brownian particle immersed in a thermal bath and subjected to a spatially periodic potential $V(x)$, which is switched on and off by a feedback external control that operates at regular intervals $\Delta t_m$. It is a feedback external control because it uses the position of the particle to switch on or off the potential, specifically it measures the position of the particle at $t_k=k\Delta t_m$ and the potential is on (off) in the time interval $J_k=(t_k,t_{k+1})$ if the instantaneous force $-V'(x(t_k))$ is positive (negative). In this way, the feedback control generates a flux to the right, even in the presence of a external force $F_{\ext}$---in a certain range of forces, $0<F_{\ext}<F_{\sto}$, at $F_{\ext}=F_{\sto}$ the flux vanishes. We have analyzed the behavior of such entropy reduction per measurement $\dSinfoPM$ as a function of the system parameters: the time interval between control updates $\Delta t_m$, the potential asymmetry $a$, the potential height $V_0$, and the applied external force $F_{\ext}$.

The entropy reduction by information per measurement $\dSinfoPM$ has been directly linked with the entropy $\Hseq(M)$ of a sequence $\bm{\Gamma}^M$ of $M$ consecutive control actions---in the long-time regime that the system reaches. Since the action of the control is dichotomic---switching either on or off the potential, the maximum number of control actions is $2^M$. If the control sequences are completely random, $\Hseq$ is maximum, whereas if only one of the values of the control is possible or the control sequences are strongly correlated, $\Hseq$ is minimum~\cite{cao_information_2009,He2023}. As a rule, $\dSinfoPM$ becomes larger as the size of the 
space of control sequences that is effectively swept by the system increases.

We have also looked into the efficiency $\eta$ of the feedback controlled Brownian ratchet. We have neatly shown that $\dSinfoPM$ must be incorporated into the thermodynamic balance of the system, to get physically reasonable values for $\eta$. In fact, the open-loop efficiency, which is simply the ratio between the output and input powers, becomes larger than unity in a certain range of parameters. Instead, the actual efficiency that has been derived by making use of the second principle, taking into account the entropy reduction of the particle $\dSinfoPM$ due to the information gathered by the control, is always smaller than unity. 

Our analytical characterization of the evolution of the joint probability of the position of the particle and the control history at the control updates have allowed us to show that the entropy of the particle plus the control is continuous thereat. On the one hand, this provides an appealing physical picture for the entropy reduction of the particle derived here from the controller plus particle perspective---which was previously derived considering the controller as an external agent~\cite{cao_thermodynamics_2009}. On the other hand, it also opens  avenues of further research. For instance, a careful choice of the relevant variables of the stochastic evolution of the feedback controlled ratchet should make it possible to write an evolution equation, valid for all times, in the ensemble picture---i.e., for a probability distribution function. In this framework, pertinent questions emerge: e..g., the stability of the long-time behavior, which could be tackled by looking for an H-theorem~\cite{van_kampen_stochastic_1992,brey_normal_1993,marconi_about_2013,garcia_de_soria_towards_2015,plata_global_2017}, or the refinement of the second principle by splitting heat into housekeeping and excess contributions~\cite{oono_steady_1998,hatano_steady-state_2001,speck_integral_2005,speck_restoring_2006,van_den_broeck_three_2010}, which could lead to a refinement of the analysis of the efficiency. 

Other interesting perspective is the impact of the protocol for the feedback controller on the observed behavior. In this paper, we have considered that the controller measures the position of the particle and updates the value of the control at regular times, a procedure that has been shown to maximize the output power delivered by the one-particle ratchet---thus the name of maximization protocol employed in the literature~\cite{cao_feedback_2004,feito_optimal_2009}. It is worth looking at other protocols that may maximize the efficiency of the one-particle ratchet, and also to investigate systems with more than one particle. For the latter case, protocols maximizing the speed of the center of mass for an infinite number of particles~\cite{feito_optimal_2009} and power for two particles in a sinusoidal potential~\cite{roca_optimal_2014} have been considered. Interestingly, the addition of a periodical oscillating force (rocking force) has been found to increase speed performance, even for one particle, increasing additionally the coherence and reproducibility~\cite{Feito2009a,Jarillo2018b}.
However, designing the protocol that maximizes power for a set of $N$ particles remains an open problem. 

Trying to formulate the open problems above in a rigorous way with the tools of optimal control theory~\cite{pontryagin_mathematical_1987,guery-odelin_driving_2023,chennakesavalu_unified_2023,blaber_optimal_2022} is also an appealing prospect, which should help improve our current understanding of how to optimize the use of the information gathered by the control. Also, improving our understanding of the thermodynamics of feedback control systems in general---and, in particular, of feedback ratchets as model systems---would contribute to the comprehension of other related problems, as the role played by information in feedback cooling~\cite{Ehrich2022} and in active matter interactions~\cite{VanSaders2023}.  

\begin{acknowledgments}
NRP and AP acknowledge financial support from Grant No.~PID2021-122588NB-I00 funded by MCIN/AEI/10.13039/501100011033 and by ``ERDF A way of making Europe.'', and also from Grant No.~ProyExcel\_00796 funded by Junta de Andalucía's PAIDI 2020 program. NRP also acknowledges support from the FPU  through Grant FPU2021/01764. NRP and AP thank Carlos A.~Plata for really useful discussions. DVM acknowledges the MCIN/AEI/10.13039/501100011033 for financial support through Grant PRE2019-088226 co-funded by the European Social Fund (ESF). FJCG acknowledges financial support from Grant No.~RTI2018-095802-B-I00 funded by the MINECO (Spain) and the European Regional Development Fund (ERDF). FJCG thanks for the kind invitation to Universidad de Sevilla during this collaboration. 
\end{acknowledgments}

\appendix

\section{Continuity of the entropy at the control
  updates}\label{sec:app-S-cont}

In this Appendix, we show that the entropy of the particle plus the
control is continuous at the control updates. Since the control has perfect knowledge of the position of
the particle and acts without error, we can construct
$P(x,\bmCk,t_{k}^{+})$ in terms of $P(x,\bmCk,t_{k}^{-})$.

The control takes two possible values after measuring the position of
the particle at time $t_{k}$: $0$ if $x\in\calX_{0}$, and $1$ if
$x\in\calX_{1}$---we do not need to specify the exact shape of the sets
$\calX_{0}$ and $\calX_{1}$, only that
$\calX_{0}\cup\calX_{1}=(-\infty,+\infty)$. We can build
$P(x,\bmCk,t_{k}^{+})$ by thinking in the frequentist way about
probabilities: let us define two functions $
\Theta_{i}(x)$ such that
$\Theta_{i}(x)=1$ if $x\in\calX_{i}$, and zero otherwise. Therefore,
\begin{subequations}\label{eq:P-tk-+}
  \begin{align}
    P(x,\bmCk=\{0,\bmCkant\},t_{k}^{+})=
    P(x,\bmCkant,t_{k}^{-})\,\Theta_{0}(x), \\
    P(x,\bmCk=\{1,\bmCkant\},t_{k}^{+})=
    P(x,\bmCkant,t_{k}^{-})\,\Theta_{1}(x).
\end{align}
\end{subequations}
Note that (i) $P(x,\bmCk,t_{k}^{+})$ is normalized as a consequence
of $P(x,\bmCkant,t_{k}^{-})$ being normalized and the fact that
$\Theta_{0}(x)+\Theta_{1}(x)\equiv 1$, and (ii) \eqref{eq:P-tk-+} can
also be interpreted as a backward equation, which univocally determines
$P(x,\bmCkant,t_{k}^{-})$ from the knowledge of
$P(x,\bmCk,t_{k}^{+})$~\footnote{Intuitively, this means that
  there is no loss of information, which hints at the continuity of
  the entropy that is proved here mathematically.}.

Now we define two associated operators
$\hat{\Theta}_{0}$ and $\hat{\Theta}_{1}$
\begin{equation}
  \label{eq:theta-C-op}
  \hat{\Theta}_{C}f(x)=f(x)\Theta_{C}(x), \quad \forall f(x);
\end{equation}
the operator $\hat{\Theta}_{C}$ restricts the function $f(x)$ to the interval $\calX_{C}$, ``cutting'' the reminder of the function. These operators are clearly linear and verify $\hat{\Theta}_{C}^{2}=\hat{\Theta}_{C}$, $\sum_{C}\hat{\Theta}_{C}=I$. In terms of them, Eq.~\eqref{eq:P-tk-+} can be written as
\begin{equation}
  \label{eq:P-tk-+-v2}
  P(x,\bmCk,t_{k}^{+})=
    \hat{\Theta}_{C_{k}} P(x,\bmCkant,t_{k}^{-}).
\end{equation}

\begin{widetext}
Let us calculate the entropy at time $t_{k}^{+}$:
\begin{align}
  S^{(k)}(t_{k}^{+})=&-k_{B}\sum_{\bmCk}\int dx  P(x,\bmCk,t_{k}^{+}) \ln
                 P(x,\bmCk,t_{k}^{+}) \nonumber \\
  =&-k_{B}\sum_{\bmCkant}\int dx  P(x,\bmCk=\{0,\bmCkant\},t_{k}^{+}) \ln
                P(x,\bmCk=\{0,\bmCkant\},t_{k}^{+}) \nonumber
  \\
  &-k_{B}\sum_{\bmCkant}\int dx  P(x,\bmCk=\{1,\bmCkant\},t_{k}^{+}) \ln
                P(x,\bmCk=\{1,\bmCkant\},t_{k}^{+}) \nonumber
  \\
  =&-k_{B} \sum_{\bmCkant}\int_{x\in\calX_{0}} dx
  P(x,\bmCkant,t_{k}^{-}) \ln P(x,\bmCkant,t_{k}^{-})
-k_{B} \sum_{\bmCkant}\int_{x\in\calX_{1}} dx
  P(x,\bmCkant,t_{k}^{-}) \ln P(x,\bmCkant,t_{k}^{-}).
                \label{eq:S-tk-+}
\end{align}
Since $\calX_{0}\cup\calX_{1}=(-\infty,+\infty)$, we conclude that
\begin{align}
  S^{(k)}(t_{k}^{+})=-k_{B} \sum_{\bmCkant}\int dx
  P(x,\bmCkant,t_{k}^{-}) \ln P(x,\bmCkant,t_{k}^{-})=S^{(k-1)}(t_{k}^{-}),
                \label{eq:S-tk-+=tk--}
\end{align}
i.e., the entropy of the particle plus the control is continuous at
each control update. 
\end{widetext}

\section{Memory loss and constancy of $\dHs$}\label{sec:app-dHs-const}

In the main text, we have shown that Eq.~\eqref{eq:M0-def} entails that the $\dHs(M)$ is constant for $M\ge M_0$. In this Appendix, we show that assuming a constant value for $\dHs(M)$ for $M\ge M_0$ implies Eq.~\eqref{eq:M0-def}. In other words, a constant value of $\dHs(M)$ for $M\ge M_0$ implies that the memory of control values is restricted to the $M_0$ most recent ones.

Let us write $\dHs(M)$ and $\dHs(M+1)$ explicitly,
\begin{subequations}
    \begin{align}
        \dHs(M)=&\sum_{\bm{C}_{k-M}^{k}} P_{\sts}(\bm{C}_{k-M}^{k}) \ln P_{\sts}(C_k|\bm{C}_{k-M}^{k-1}), \\
        \dHs(M+1)=&\sum_{\bm{C}_{k-M-1}^{k}} P_{\sts}(\bm{C}_{k-M-1}^{k}) \ln P_{\sts}(C_k|\bm{C}_{k-M-1}^{k-1});
    \end{align}
\end{subequations}
we recall that 
\begin{subequations}
    \begin{align}
        \bm{C}_{k-M}^{k-1}=&(C_{k-1},\ldots,C_{k-M}), \\
        \bm{C}_{k-M-1}^{k-1}=&(C_{k-1},\ldots,C_{k-M},C_{k-M-1}).
    \end{align}
\end{subequations}
involve $M$ and $M+1$ control values, respectively. 

By employing the Kullblack-Leibler divergence of $P_{\sts}(C_{k},C_{k-M-1}|\bm{C}_{k-M}^{k-1})$ with respect to the factorized distribution $P_{\sts}(C_{k}|\bm{C}_{k-M}^{k-1})P_{\sts}(C_{k-M-1}|\bm{C}_{k-M}^{k-1})$, it can be shown that~\cite{cover_elements_2006}
\begin{equation}
    \dHs(M+1)\le \dHs(M), 
\end{equation}
with the equality holding if and only if the conditional probability $P_{\sts}(C_{k},C_{k-M-1}|\bm{C}_{k-M}^{k-1})$ factorizes, i.e., if and only if
\begin{align}
    P_{\sts}(C_{k},C_{k-M-1}&|\bm{C}_{k-M}^{k-1}) \nonumber \\
    &=P_{\sts}(C_{k}|\bm{C}_{k-M}^{k-1})P_{\sts}(C_{k-M-1}|\bm{C}_{k-M}^{k-1}).
    \label{eq:app-nec-suff}
\end{align}
Employing Bayes's theorem, specifically
\begin{align}
    P_{\sts}(C_{k-M-1}|\bm{C}_{k-M}^{k-1})=\frac{P_{\sts}(\bm{C}_{k-M-1}^{k-1})}{P_{\sts}(\bm{C}_{k-M}^{k-1})},
\end{align}
Eq.~\eqref{eq:app-nec-suff} is equivalent to
\begin{align}
    P_{\sts}(\bm{C}_{k-M-1}^k)=P_{\sts}(C_{k}|\bm{C}_{k-M}^{k-1})P_{\sts}(\bm{C}_{k-M-1}^{k-1}),
\end{align}
i.e., equivalent to 
\begin{align}
    P_{\sts}(C_k|\bm{C}_{k-M-1}^{k-1})=P_{\sts}(C_{k}|\bm{C}_{k-M}^{k-1}),
    \label{eq:app-nec-suff-result}
\end{align}
making use again of the Bayes theorem.

Equation~\eqref{eq:app-nec-suff-result} is the desired result: incorporating the extra condition $C_{k-M-1}$ on the lhs do not change the probability, physically this means that the memory of control values is lost after the considered $M$ values $\bm{C}_{k-M}^{k-1}=(C_{k-1},\ldots,C_{k-M})$ on the rhs. That is, if $\dHs(M)$ is constant for $M\ge M_0$, we can use Eq.~\eqref{eq:app-nec-suff-result} recursively to write
\begin{align}
   P_{\sts}(C_{k}|\bm{C}_{k-M_0}^{k-1})&=P_{\sts}(C_k|\bm{C}_{k-M_0-1}^{k-1}) \nonumber \\
   &=P_{\sts}(C_k|\bm{C}_{k-M_0-2}^{k-1})=\cdots,
\end{align}
i.e.
\begin{equation}
    P_{\sts}(C_{k}|\bm{C}_{k-M}^{k-1})=P_{\sts}(C_{k}|\bm{C}_{k-M_0}^{k-1}), \quad \forall M\ge M_0,  
\end{equation}
which is precisely Eq.~\eqref{eq:M0-def}.


%


\begin{thebibliography}{71}%
\makeatletter
\providecommand \@ifxundefined [1]{%
 \@ifx{#1\undefined}
}%
\providecommand \@ifnum [1]{%
 \ifnum #1\expandafter \@firstoftwo
 \else \expandafter \@secondoftwo
 \fi
}%
\providecommand \@ifx [1]{%
 \ifx #1\expandafter \@firstoftwo
 \else \expandafter \@secondoftwo
 \fi
}%
\providecommand \natexlab [1]{#1}%
\providecommand \enquote  [1]{``#1''}%
\providecommand \bibnamefont  [1]{#1}%
\providecommand \bibfnamefont [1]{#1}%
\providecommand \citenamefont [1]{#1}%
\providecommand \href@noop [0]{\@secondoftwo}%
\providecommand \href [0]{\begingroup \@sanitize@url \@href}%
\providecommand \@href[1]{\@@startlink{#1}\@@href}%
\providecommand \@@href[1]{\endgroup#1\@@endlink}%
\providecommand \@sanitize@url [0]{\catcode `\\12\catcode `\$12\catcode
  `\&12\catcode `\#12\catcode `\^12\catcode `\_12\catcode `\%12\relax}%
\providecommand \@@startlink[1]{}%
\providecommand \@@endlink[0]{}%
\providecommand \url  [0]{\begingroup\@sanitize@url \@url }%
\providecommand \@url [1]{\endgroup\@href {#1}{\urlprefix }}%
\providecommand \urlprefix  [0]{URL }%
\providecommand \Eprint [0]{\href }%
\providecommand \doibase [0]{https://doi.org/}%
\providecommand \selectlanguage [0]{\@gobble}%
\providecommand \bibinfo  [0]{\@secondoftwo}%
\providecommand \bibfield  [0]{\@secondoftwo}%
\providecommand \translation [1]{[#1]}%
\providecommand \BibitemOpen [0]{}%
\providecommand \bibitemStop [0]{}%
\providecommand \bibitemNoStop [0]{.\EOS\space}%
\providecommand \EOS [0]{\spacefactor3000\relax}%
\providecommand \BibitemShut  [1]{\csname bibitem#1\endcsname}%
\let\auto@bib@innerbib\@empty
\bibitem [{\citenamefont {Bechhoefer}(2005)}]{bechhoefer_feedback_2005}%
  \BibitemOpen
  \bibfield  {author} {\bibinfo {author} {\bibfnamefont {J.}~\bibnamefont
  {Bechhoefer}},\ }\bibfield  {title} {\bibinfo {title} {Feedback for
  physicists: {A} tutorial essay on control},\ }\href
  {https://doi.org/10.1103/RevModPhys.77.783} {\bibfield  {journal} {\bibinfo
  {journal} {Reviews of Modern Physics}\ }\textbf {\bibinfo {volume} {77}},\
  \bibinfo {pages} {783} (\bibinfo {year} {2005})}\BibitemShut {NoStop}%
\bibitem [{\citenamefont {Leff}\ and\ \citenamefont {Rex}(2002)}]{Leff2002}%
  \BibitemOpen
  \bibinfo {editor} {\bibfnamefont {H.}~\bibnamefont {Leff}}\ and\ \bibinfo
  {editor} {\bibfnamefont {A.~F.}\ \bibnamefont {Rex}},\ eds.,\ \href@noop {}
  {\emph {\bibinfo {title} {{Maxwell's Demon 2 Entropy, Classical and Quantum
  Information, Computing}}}}\ (\bibinfo  {publisher} {CRC Press},\ \bibinfo
  {year} {2002})\ p.\ \bibinfo {pages} {502}\BibitemShut {NoStop}%
\bibitem [{\citenamefont {Szilard}(1929)}]{Szilard1929}%
  \BibitemOpen
  \bibfield  {author} {\bibinfo {author} {\bibfnamefont {L.}~\bibnamefont
  {Szilard}},\ }\bibfield  {title} {\bibinfo {title} {{{\"{u}}ber die
  Entropieverminderung in einem thermodynamischen System bei Eingriffen
  intelligenter Wesen}},\ }\href {https://doi.org/10.1007/BF01341281}
  {\bibfield  {journal} {\bibinfo  {journal} {Zeitschrift f{\"{u}}r Physik}\
  }\textbf {\bibinfo {volume} {53}},\ \bibinfo {pages} {840} (\bibinfo {year}
  {1929})}\BibitemShut {NoStop}%
\bibitem [{\citenamefont {Szilard}(1964)}]{Szilard1964}%
  \BibitemOpen
  \bibfield  {author} {\bibinfo {author} {\bibfnamefont {L.}~\bibnamefont
  {Szilard}},\ }\bibfield  {title} {\bibinfo {title} {{On the decrease of
  entropy in a thermodynamic system by the intervention of intelligent
  beings}},\ }\href {https://doi.org/10.1002/bs.3830090402} {\bibfield
  {journal} {\bibinfo  {journal} {Behavioral Science}\ }\textbf {\bibinfo
  {volume} {9}},\ \bibinfo {pages} {301} (\bibinfo {year} {1964})}\BibitemShut
  {NoStop}%
\bibitem [{\citenamefont {Shannon}\ and\ \citenamefont
  {Weaver}(1949)}]{Shannon1949}%
  \BibitemOpen
  \bibfield  {author} {\bibinfo {author} {\bibfnamefont {C.}~\bibnamefont
  {Shannon}}\ and\ \bibinfo {author} {\bibfnamefont {W.}~\bibnamefont
  {Weaver}},\ }\href@noop {} {\emph {\bibinfo {title} {{The Mathematical Theory
  of Communication}}}}\ (\bibinfo  {publisher} {University of Illinois Press},\
  \bibinfo {year} {1949})\BibitemShut {NoStop}%
\bibitem [{\citenamefont {Cover}\ and\ \citenamefont
  {Thomas}(2006)}]{cover_elements_2006}%
  \BibitemOpen
  \bibfield  {author} {\bibinfo {author} {\bibfnamefont {T.~M.}\ \bibnamefont
  {Cover}}\ and\ \bibinfo {author} {\bibfnamefont {J.~A.}\ \bibnamefont
  {Thomas}},\ }\href@noop {} {\emph {\bibinfo {title} {Elements of
  {Information} {Theory}}}},\ \bibinfo {edition} {2nd}\ ed.\ (\bibinfo
  {publisher} {Wiley-Interscience},\ \bibinfo {address} {Hoboken, New Jersey},\
  \bibinfo {year} {2006})\BibitemShut {NoStop}%
\bibitem [{\citenamefont {Landauer}(1961)}]{Landauer1961}%
  \BibitemOpen
  \bibfield  {author} {\bibinfo {author} {\bibfnamefont {R.}~\bibnamefont
  {Landauer}},\ }\bibfield  {title} {\bibinfo {title} {{Irreversibility and
  Heat Generation in the Computing Process}},\ }\href
  {https://doi.org/10.1147/rd.53.0183} {\bibfield  {journal} {\bibinfo
  {journal} {IBM Journal of Research and Development}\ }\textbf {\bibinfo
  {volume} {5}},\ \bibinfo {pages} {183} (\bibinfo {year} {1961})}\BibitemShut
  {NoStop}%
\bibitem [{\citenamefont {Bennett}(1982)}]{Bennett1982}%
  \BibitemOpen
  \bibfield  {author} {\bibinfo {author} {\bibfnamefont {C.~H.}\ \bibnamefont
  {Bennett}},\ }\bibfield  {title} {\bibinfo {title} {{The thermodynamics of
  computation-a review}},\ }\href {https://doi.org/10.1007/BF02084158}
  {\bibfield  {journal} {\bibinfo  {journal} {International Journal of
  Theoretical Physics}\ }\textbf {\bibinfo {volume} {21}},\ \bibinfo {pages}
  {905} (\bibinfo {year} {1982})}\BibitemShut {NoStop}%
\bibitem [{\citenamefont {Touchette}\ and\ \citenamefont
  {Lloyd}(2000)}]{Touchette2000}%
  \BibitemOpen
  \bibfield  {author} {\bibinfo {author} {\bibfnamefont {H.}~\bibnamefont
  {Touchette}}\ and\ \bibinfo {author} {\bibfnamefont {S.}~\bibnamefont
  {Lloyd}},\ }\bibfield  {title} {\bibinfo {title} {{Information-theoretic
  limits of control}},\ }\href {http://www.ncbi.nlm.nih.gov/pubmed/11017467}
  {\bibfield  {journal} {\bibinfo  {journal} {Physical review letters}\
  }\textbf {\bibinfo {volume} {84}},\ \bibinfo {pages} {1156} (\bibinfo {year}
  {2000})}\BibitemShut {NoStop}%
\bibitem [{\citenamefont {Touchette}\ and\ \citenamefont
  {Lloyd}(2004)}]{Touchette2004}%
  \BibitemOpen
  \bibfield  {author} {\bibinfo {author} {\bibfnamefont {H.}~\bibnamefont
  {Touchette}}\ and\ \bibinfo {author} {\bibfnamefont {S.}~\bibnamefont
  {Lloyd}},\ }\bibfield  {title} {\bibinfo {title} {{Information-theoretic
  approach to the study of control systems}},\ }\href
  {https://doi.org/10.1016/j.physa.2003.09.007} {\bibfield  {journal} {\bibinfo
   {journal} {Physica A: Statistical Mechanics and its Applications}\ }\textbf
  {\bibinfo {volume} {331}},\ \bibinfo {pages} {140} (\bibinfo {year}
  {2004})}\BibitemShut {NoStop}%
\bibitem [{\citenamefont {Quan}\ \emph {et~al.}(2006)\citenamefont {Quan},
  \citenamefont {Wang}, \citenamefont {Liu}, \citenamefont {Sun},\ and\
  \citenamefont {Nori}}]{Quan2006}%
  \BibitemOpen
  \bibfield  {author} {\bibinfo {author} {\bibfnamefont {H.~T.}\ \bibnamefont
  {Quan}}, \bibinfo {author} {\bibfnamefont {Y.~D.}\ \bibnamefont {Wang}},
  \bibinfo {author} {\bibfnamefont {Y.-x.}\ \bibnamefont {Liu}}, \bibinfo
  {author} {\bibfnamefont {C.~P.}\ \bibnamefont {Sun}},\ and\ \bibinfo {author}
  {\bibfnamefont {F.}~\bibnamefont {Nori}},\ }\bibfield  {title} {\bibinfo
  {title} {{Maxwell's Demon Assisted Thermodynamic Cycle in Superconducting
  Quantum Circuits}},\ }\href {https://doi.org/10.1103/PhysRevLett.97.180402}
  {\bibfield  {journal} {\bibinfo  {journal} {Physical Review Letters}\
  }\textbf {\bibinfo {volume} {97}},\ \bibinfo {pages} {180402} (\bibinfo
  {year} {2006})}\BibitemShut {NoStop}%
\bibitem [{\citenamefont {Sagawa}\ and\ \citenamefont
  {Ueda}(2008)}]{Sagawa2008}%
  \BibitemOpen
  \bibfield  {author} {\bibinfo {author} {\bibfnamefont {T.}~\bibnamefont
  {Sagawa}}\ and\ \bibinfo {author} {\bibfnamefont {M.}~\bibnamefont {Ueda}},\
  }\bibfield  {title} {\bibinfo {title} {{Second Law of Thermodynamics with
  Discrete Quantum Feedback Control}},\ }\href
  {https://doi.org/10.1103/PhysRevLett.100.080403} {\bibfield  {journal}
  {\bibinfo  {journal} {Physical Review Letters}\ }\textbf {\bibinfo {volume}
  {100}},\ \bibinfo {pages} {080403} (\bibinfo {year} {2008})}\BibitemShut
  {NoStop}%
\bibitem [{\citenamefont {Cao}\ \emph {et~al.}(2009)\citenamefont {Cao},
  \citenamefont {Feito},\ and\ \citenamefont
  {Touchette}}]{cao_information_2009}%
  \BibitemOpen
  \bibfield  {author} {\bibinfo {author} {\bibfnamefont {F.}~\bibnamefont
  {Cao}}, \bibinfo {author} {\bibfnamefont {M.}~\bibnamefont {Feito}},\ and\
  \bibinfo {author} {\bibfnamefont {H.}~\bibnamefont {Touchette}},\ }\bibfield
  {title} {\bibinfo {title} {Information and flux in a feedback controlled
  {Brownian} ratchet},\ }\href {https://doi.org/10.1016/j.physa.2008.10.006}
  {\bibfield  {journal} {\bibinfo  {journal} {Physica A: Statistical Mechanics
  and its Applications}\ }\textbf {\bibinfo {volume} {388}},\ \bibinfo {pages}
  {113} (\bibinfo {year} {2009})}\BibitemShut {NoStop}%
\bibitem [{\citenamefont {Cao}\ and\ \citenamefont
  {Feito}(2009)}]{cao_thermodynamics_2009}%
  \BibitemOpen
  \bibfield  {author} {\bibinfo {author} {\bibfnamefont {F.~J.}\ \bibnamefont
  {Cao}}\ and\ \bibinfo {author} {\bibfnamefont {M.}~\bibnamefont {Feito}},\
  }\bibfield  {title} {\bibinfo {title} {Thermodynamics of feedback controlled
  systems},\ }\href {https://doi.org/10.1103/PhysRevE.79.041118} {\bibfield
  {journal} {\bibinfo  {journal} {Physical Review E}\ }\textbf {\bibinfo
  {volume} {79}},\ \bibinfo {pages} {041118} (\bibinfo {year}
  {2009})}\BibitemShut {NoStop}%
\bibitem [{\citenamefont {Mandal}\ and\ \citenamefont
  {Jarzynski}(2012)}]{Mandal2012}%
  \BibitemOpen
  \bibfield  {author} {\bibinfo {author} {\bibfnamefont {D.}~\bibnamefont
  {Mandal}}\ and\ \bibinfo {author} {\bibfnamefont {C.}~\bibnamefont
  {Jarzynski}},\ }\bibfield  {title} {\bibinfo {title} {{Work and information
  processing in a solvable model of Maxwell's demon}},\ }\href
  {https://doi.org/10.1073/pnas.1204263109} {\bibfield  {journal} {\bibinfo
  {journal} {Proceedings of the National Academy of Sciences}\ }\textbf
  {\bibinfo {volume} {109}},\ \bibinfo {pages} {11641} (\bibinfo {year}
  {2012})}\BibitemShut {NoStop}%
\bibitem [{\citenamefont {Barato}\ and\ \citenamefont
  {Seifert}(2014)}]{Barato2014b}%
  \BibitemOpen
  \bibfield  {author} {\bibinfo {author} {\bibfnamefont {A.~C.}\ \bibnamefont
  {Barato}}\ and\ \bibinfo {author} {\bibfnamefont {U.}~\bibnamefont
  {Seifert}},\ }\bibfield  {title} {\bibinfo {title} {{Unifying Three
  Perspectives on Information Processing in Stochastic Thermodynamics}},\
  }\href {https://doi.org/10.1103/PhysRevLett.112.090601} {\bibfield  {journal}
  {\bibinfo  {journal} {Physical Review Letters}\ }\textbf {\bibinfo {volume}
  {112}},\ \bibinfo {pages} {090601} (\bibinfo {year} {2014})}\BibitemShut
  {NoStop}%
\bibitem [{\citenamefont {Van~Vu}\ and\ \citenamefont
  {Hasegawa}(2020)}]{van_vu_uncertainty_2020}%
  \BibitemOpen
  \bibfield  {author} {\bibinfo {author} {\bibfnamefont {T.}~\bibnamefont
  {Van~Vu}}\ and\ \bibinfo {author} {\bibfnamefont {Y.}~\bibnamefont
  {Hasegawa}},\ }\bibfield  {title} {\bibinfo {title} {Uncertainty relation
  under information measurement and feedback control},\ }\href
  {https://doi.org/10.1088/1751-8121/ab64a4} {\bibfield  {journal} {\bibinfo
  {journal} {Journal of Physics A: Mathematical and Theoretical}\ }\textbf
  {\bibinfo {volume} {53}},\ \bibinfo {pages} {075001} (\bibinfo {year}
  {2020})}\BibitemShut {NoStop}%
\bibitem [{\citenamefont {Lozano-Durán}\ and\ \citenamefont
  {Arranz}(2022)}]{lozano-duran_information-theoretic_2022}%
  \BibitemOpen
  \bibfield  {author} {\bibinfo {author} {\bibfnamefont {A.}~\bibnamefont
  {Lozano-Durán}}\ and\ \bibinfo {author} {\bibfnamefont {G.}~\bibnamefont
  {Arranz}},\ }\bibfield  {title} {\bibinfo {title} {Information-theoretic
  formulation of dynamical systems: {Causality}, modeling, and control},\
  }\href {https://doi.org/10.1103/PhysRevResearch.4.023195} {\bibfield
  {journal} {\bibinfo  {journal} {Physical Review Research}\ }\textbf {\bibinfo
  {volume} {4}},\ \bibinfo {pages} {023195} (\bibinfo {year}
  {2022})}\BibitemShut {NoStop}%
\bibitem [{\citenamefont {Bhattacharyya}\ and\ \citenamefont
  {Jarzynski}(2022)}]{bhattacharyya_feedback-controlled_2022}%
  \BibitemOpen
  \bibfield  {author} {\bibinfo {author} {\bibfnamefont {D.}~\bibnamefont
  {Bhattacharyya}}\ and\ \bibinfo {author} {\bibfnamefont {C.}~\bibnamefont
  {Jarzynski}},\ }\bibfield  {title} {\bibinfo {title} {From a
  feedback-controlled demon to an information ratchet in a double quantum
  dot},\ }\href {https://doi.org/10.1103/PhysRevE.106.064101} {\bibfield
  {journal} {\bibinfo  {journal} {Physical Review E}\ }\textbf {\bibinfo
  {volume} {106}},\ \bibinfo {pages} {064101} (\bibinfo {year}
  {2022})}\BibitemShut {NoStop}%
\bibitem [{\citenamefont {Cao}\ \emph {et~al.}(2004)\citenamefont {Cao},
  \citenamefont {Dinis},\ and\ \citenamefont {Parrondo}}]{cao_feedback_2004}%
  \BibitemOpen
  \bibfield  {author} {\bibinfo {author} {\bibfnamefont {F.~J.}\ \bibnamefont
  {Cao}}, \bibinfo {author} {\bibfnamefont {L.}~\bibnamefont {Dinis}},\ and\
  \bibinfo {author} {\bibfnamefont {J.~M.~R.}\ \bibnamefont {Parrondo}},\
  }\bibfield  {title} {\bibinfo {title} {Feedback {Control} in a {Collective}
  {Flashing} {Ratchet}},\ }\href
  {https://doi.org/10.1103/PhysRevLett.93.040603} {\bibfield  {journal}
  {\bibinfo  {journal} {Physical Review Letters}\ }\textbf {\bibinfo {volume}
  {93}},\ \bibinfo {pages} {040603} (\bibinfo {year} {2004})}\BibitemShut
  {NoStop}%
\bibitem [{\citenamefont {Abreu}\ and\ \citenamefont
  {Seifert}(2012)}]{abreu_thermodynamics_2012}%
  \BibitemOpen
  \bibfield  {author} {\bibinfo {author} {\bibfnamefont {D.}~\bibnamefont
  {Abreu}}\ and\ \bibinfo {author} {\bibfnamefont {U.}~\bibnamefont
  {Seifert}},\ }\bibfield  {title} {\bibinfo {title} {Thermodynamics of
  {Genuine} {Nonequilibrium} {States} under {Feedback} {Control}},\ }\href
  {https://doi.org/10.1103/PhysRevLett.108.030601} {\bibfield  {journal}
  {\bibinfo  {journal} {Physical Review Letters}\ }\textbf {\bibinfo {volume}
  {108}},\ \bibinfo {pages} {030601} (\bibinfo {year} {2012})}\BibitemShut
  {NoStop}%
\bibitem [{\citenamefont {Lucero}\ \emph {et~al.}(2021)\citenamefont {Lucero},
  \citenamefont {Ehrich}, \citenamefont {Bechhoefer},\ and\ \citenamefont
  {Sivak}}]{lucero_maximal_2021}%
  \BibitemOpen
  \bibfield  {author} {\bibinfo {author} {\bibfnamefont {J.~N.~E.}\
  \bibnamefont {Lucero}}, \bibinfo {author} {\bibfnamefont {J.}~\bibnamefont
  {Ehrich}}, \bibinfo {author} {\bibfnamefont {J.}~\bibnamefont {Bechhoefer}},\
  and\ \bibinfo {author} {\bibfnamefont {D.~A.}\ \bibnamefont {Sivak}},\
  }\bibfield  {title} {\bibinfo {title} {Maximal fluctuation exploitation in
  {Gaussian} information engines},\ }\href
  {https://doi.org/10.1103/PhysRevE.104.044122} {\bibfield  {journal} {\bibinfo
   {journal} {Physical Review E}\ }\textbf {\bibinfo {volume} {104}},\ \bibinfo
  {pages} {044122} (\bibinfo {year} {2021})}\BibitemShut {NoStop}%
\bibitem [{\citenamefont {Lopez}\ \emph {et~al.}(2008)\citenamefont {Lopez},
  \citenamefont {Kuwada}, \citenamefont {Craig}, \citenamefont {Long},\ and\
  \citenamefont {Linke}}]{Lopez2008}%
  \BibitemOpen
  \bibfield  {author} {\bibinfo {author} {\bibfnamefont {B.~J.~B.}\
  \bibnamefont {Lopez}}, \bibinfo {author} {\bibfnamefont {N.~J.~N.}\
  \bibnamefont {Kuwada}}, \bibinfo {author} {\bibfnamefont {E.~E.~M.}\
  \bibnamefont {Craig}}, \bibinfo {author} {\bibfnamefont {B.~R.}\ \bibnamefont
  {Long}},\ and\ \bibinfo {author} {\bibfnamefont {H.}~\bibnamefont {Linke}},\
  }\bibfield  {title} {\bibinfo {title} {{Realization of a Feedback Controlled
  Flashing Ratchet}},\ }\href {https://doi.org/10.1103/PhysRevLett.101.220601}
  {\bibfield  {journal} {\bibinfo  {journal} {Physical Review Letters}\
  }\textbf {\bibinfo {volume} {101}},\ \bibinfo {pages} {220601} (\bibinfo
  {year} {2008})}\BibitemShut {NoStop}%
\bibitem [{\citenamefont {Toyabe}\ \emph {et~al.}(2010)\citenamefont {Toyabe},
  \citenamefont {Sagawa}, \citenamefont {Ueda}, \citenamefont {Muneyuki},\ and\
  \citenamefont {Sano}}]{Toyabe2010}%
  \BibitemOpen
  \bibfield  {author} {\bibinfo {author} {\bibfnamefont {S.}~\bibnamefont
  {Toyabe}}, \bibinfo {author} {\bibfnamefont {T.}~\bibnamefont {Sagawa}},
  \bibinfo {author} {\bibfnamefont {M.}~\bibnamefont {Ueda}}, \bibinfo {author}
  {\bibfnamefont {E.}~\bibnamefont {Muneyuki}},\ and\ \bibinfo {author}
  {\bibfnamefont {M.}~\bibnamefont {Sano}},\ }\bibfield  {title} {\bibinfo
  {title} {{Experimental demonstration of information-to-energy conversion and
  validation of the generalized Jarzynski equality}},\ }\href
  {https://doi.org/10.1038/nphys1821} {\bibfield  {journal} {\bibinfo
  {journal} {Nature Physics}\ }\textbf {\bibinfo {volume} {6}},\ \bibinfo
  {pages} {988} (\bibinfo {year} {2010})}\BibitemShut {NoStop}%
\bibitem [{\citenamefont {Debiossac}\ \emph {et~al.}(2020)\citenamefont
  {Debiossac}, \citenamefont {Grass}, \citenamefont {Alonso}, \citenamefont
  {Lutz},\ and\ \citenamefont {Kiesel}}]{Debiossac2020}%
  \BibitemOpen
  \bibfield  {author} {\bibinfo {author} {\bibfnamefont {M.}~\bibnamefont
  {Debiossac}}, \bibinfo {author} {\bibfnamefont {D.}~\bibnamefont {Grass}},
  \bibinfo {author} {\bibfnamefont {J.~J.}\ \bibnamefont {Alonso}}, \bibinfo
  {author} {\bibfnamefont {E.}~\bibnamefont {Lutz}},\ and\ \bibinfo {author}
  {\bibfnamefont {N.}~\bibnamefont {Kiesel}},\ }\bibfield  {title} {\bibinfo
  {title} {{Thermodynamics of continuous non-Markovian feedback control}},\
  }\href {https://doi.org/10.1038/s41467-020-15148-5} {\bibfield  {journal}
  {\bibinfo  {journal} {Nature Communications 2020 11:1}\ }\textbf {\bibinfo
  {volume} {11}},\ \bibinfo {pages} {1} (\bibinfo {year} {2020})},\ \Eprint
  {https://arxiv.org/abs/1904.04889} {1904.04889} \BibitemShut {NoStop}%
\bibitem [{\citenamefont {Saha}\ \emph {et~al.}(2022)\citenamefont {Saha},
  \citenamefont {Lucero}, \citenamefont {Ehrich}, \citenamefont {Sivak},\ and\
  \citenamefont {Bechhoefer}}]{saha_bayesian_2022}%
  \BibitemOpen
  \bibfield  {author} {\bibinfo {author} {\bibfnamefont {T.~K.}\ \bibnamefont
  {Saha}}, \bibinfo {author} {\bibfnamefont {J.~N.}\ \bibnamefont {Lucero}},
  \bibinfo {author} {\bibfnamefont {J.}~\bibnamefont {Ehrich}}, \bibinfo
  {author} {\bibfnamefont {D.~A.}\ \bibnamefont {Sivak}},\ and\ \bibinfo
  {author} {\bibfnamefont {J.}~\bibnamefont {Bechhoefer}},\ }\bibfield  {title}
  {\bibinfo {title} {Bayesian {Information} {Engine} that {Optimally}
  {Exploits} {Noisy} {Measurements}},\ }\href
  {https://doi.org/10.1103/PhysRevLett.129.130601} {\bibfield  {journal}
  {\bibinfo  {journal} {Physical Review Letters}\ }\textbf {\bibinfo {volume}
  {129}},\ \bibinfo {pages} {130601} (\bibinfo {year} {2022})}\BibitemShut
  {NoStop}%
\bibitem [{\citenamefont {Feito}\ and\ \citenamefont
  {Cao}(2009)}]{feito_optimal_2009}%
  \BibitemOpen
  \bibfield  {author} {\bibinfo {author} {\bibfnamefont {M.}~\bibnamefont
  {Feito}}\ and\ \bibinfo {author} {\bibfnamefont {F.~J.}\ \bibnamefont
  {Cao}},\ }\bibfield  {title} {\bibinfo {title} {Optimal operation of feedback
  flashing ratchets},\ }\href
  {https://doi.org/10.1088/1742-5468/2009/01/P01031} {\bibfield  {journal}
  {\bibinfo  {journal} {Journal of Statistical Mechanics: Theory and
  Experiment}\ }\textbf {\bibinfo {volume} {2009}},\ \bibinfo {pages} {P01031}
  (\bibinfo {year} {2009})}\BibitemShut {NoStop}%
\bibitem [{\citenamefont {Barato}\ \emph {et~al.}(2014)\citenamefont {Barato},
  \citenamefont {Hartich},\ and\ \citenamefont {Seifert}}]{Barato2014h}%
  \BibitemOpen
  \bibfield  {author} {\bibinfo {author} {\bibfnamefont {A.~C.}\ \bibnamefont
  {Barato}}, \bibinfo {author} {\bibfnamefont {D.}~\bibnamefont {Hartich}},\
  and\ \bibinfo {author} {\bibfnamefont {U.}~\bibnamefont {Seifert}},\
  }\bibfield  {title} {\bibinfo {title} {{Efficiency of cellular information
  processing}},\ }\href {https://doi.org/10.1088/1367-2630/16/10/103024}
  {\bibfield  {journal} {\bibinfo  {journal} {New Journal of Physics}\ }\textbf
  {\bibinfo {volume} {16}},\ \bibinfo {pages} {103024} (\bibinfo {year}
  {2014})},\ \Eprint {https://arxiv.org/abs/1405.7241} {1405.7241} \BibitemShut
  {NoStop}%
\bibitem [{\citenamefont {Barato}\ and\ \citenamefont
  {Seifert}(2015)}]{Barato2015}%
  \BibitemOpen
  \bibfield  {author} {\bibinfo {author} {\bibfnamefont {A.~C.}\ \bibnamefont
  {Barato}}\ and\ \bibinfo {author} {\bibfnamefont {U.}~\bibnamefont
  {Seifert}},\ }\bibfield  {title} {\bibinfo {title} {{Thermodynamic
  Uncertainty Relation for Biomolecular Processes}},\ }\href
  {https://doi.org/10.1103/PhysRevLett.114.158101} {\bibfield  {journal}
  {\bibinfo  {journal} {Physical Review Letters}\ }\textbf {\bibinfo {volume}
  {114}},\ \bibinfo {pages} {158101} (\bibinfo {year} {2015})}\BibitemShut
  {NoStop}%
\bibitem [{\citenamefont {Lee}\ \emph {et~al.}(2022)\citenamefont {Lee},
  \citenamefont {Lee}, \citenamefont {Kwon},\ and\ \citenamefont
  {Park}}]{lee_speed_2022}%
  \BibitemOpen
  \bibfield  {author} {\bibinfo {author} {\bibfnamefont {J.~S.}\ \bibnamefont
  {Lee}}, \bibinfo {author} {\bibfnamefont {S.}~\bibnamefont {Lee}}, \bibinfo
  {author} {\bibfnamefont {H.}~\bibnamefont {Kwon}},\ and\ \bibinfo {author}
  {\bibfnamefont {H.}~\bibnamefont {Park}},\ }\bibfield  {title} {\bibinfo
  {title} {Speed {Limit} for a {Highly} {Irreversible} {Process} and {Tight}
  {Finite}-{Time} {Landauer}’s {Bound}},\ }\href
  {https://doi.org/10.1103/PhysRevLett.129.120603} {\bibfield  {journal}
  {\bibinfo  {journal} {Physical Review Letters}\ }\textbf {\bibinfo {volume}
  {129}},\ \bibinfo {pages} {120603} (\bibinfo {year} {2022})}\BibitemShut
  {NoStop}%
\bibitem [{\citenamefont {Seifert}(2012)}]{seifert_stochastic_2012}%
  \BibitemOpen
  \bibfield  {author} {\bibinfo {author} {\bibfnamefont {U.}~\bibnamefont
  {Seifert}},\ }\bibfield  {title} {\bibinfo {title} {Stochastic
  thermodynamics, fluctuation theorems and molecular machines},\ }\href
  {https://doi.org/10.1088/0034-4885/75/12/126001} {\bibfield  {journal}
  {\bibinfo  {journal} {Reports on Progress in Physics}\ }\textbf {\bibinfo
  {volume} {75}},\ \bibinfo {pages} {126001} (\bibinfo {year}
  {2012})}\BibitemShut {NoStop}%
\bibitem [{\citenamefont {Sagawa}\ and\ \citenamefont
  {Ueda}(2013)}]{Sagawa2013a}%
  \BibitemOpen
  \bibfield  {author} {\bibinfo {author} {\bibfnamefont {T.}~\bibnamefont
  {Sagawa}}\ and\ \bibinfo {author} {\bibfnamefont {M.}~\bibnamefont {Ueda}},\
  }\bibfield  {title} {\bibinfo {title} {{Role of mutual information in entropy
  production under information exchanges}},\ }\href
  {https://doi.org/10.1088/1367-2630/15/12/125012} {\bibfield  {journal}
  {\bibinfo  {journal} {New Journal of Physics}\ }\textbf {\bibinfo {volume}
  {15}},\ \bibinfo {pages} {125012} (\bibinfo {year} {2013})},\ \Eprint
  {https://arxiv.org/abs/1307.6092} {1307.6092} \BibitemShut {NoStop}%
\bibitem [{\citenamefont {Bechhoefer}(2015)}]{Bechhoefer2015}%
  \BibitemOpen
  \bibfield  {author} {\bibinfo {author} {\bibfnamefont {J.}~\bibnamefont
  {Bechhoefer}},\ }\bibfield  {title} {\bibinfo {title} {{Hidden Markov models
  for stochastic thermodynamics}},\ }\href
  {https://doi.org/10.1088/1367-2630/17/7/075003} {\bibfield  {journal}
  {\bibinfo  {journal} {New Journal of Physics}\ }\textbf {\bibinfo {volume}
  {17}},\ \bibinfo {pages} {075003} (\bibinfo {year} {2015})},\ \Eprint
  {https://arxiv.org/abs/1504.00293} {1504.00293} \BibitemShut {NoStop}%
\bibitem [{\citenamefont {Potts}\ and\ \citenamefont
  {Samuelsson}(2018)}]{Potts2018}%
  \BibitemOpen
  \bibfield  {author} {\bibinfo {author} {\bibfnamefont {P.~P.}\ \bibnamefont
  {Potts}}\ and\ \bibinfo {author} {\bibfnamefont {P.}~\bibnamefont
  {Samuelsson}},\ }\bibfield  {title} {\bibinfo {title} {{Detailed Fluctuation
  Relation for Arbitrary Measurement and Feedback Schemes}},\ }\href
  {https://doi.org/10.1103/PhysRevLett.121.210603} {\bibfield  {journal}
  {\bibinfo  {journal} {Physical Review Letters}\ }\textbf {\bibinfo {volume}
  {121}},\ \bibinfo {pages} {210603} (\bibinfo {year} {2018})}\BibitemShut
  {NoStop}%
\bibitem [{\citenamefont {Potts}\ and\ \citenamefont
  {Samuelsson}(2019)}]{Potts2019}%
  \BibitemOpen
  \bibfield  {author} {\bibinfo {author} {\bibfnamefont {P.~P.}\ \bibnamefont
  {Potts}}\ and\ \bibinfo {author} {\bibfnamefont {P.}~\bibnamefont
  {Samuelsson}},\ }\bibfield  {title} {\bibinfo {title} {{Thermodynamic
  uncertainty relations including measurement and feedback}},\ }\href
  {https://doi.org/10.1103/PhysRevE.100.052137} {\bibfield  {journal} {\bibinfo
   {journal} {Physical Review E}\ }\textbf {\bibinfo {volume} {100}},\ \bibinfo
  {pages} {052137} (\bibinfo {year} {2019})}\BibitemShut {NoStop}%
\bibitem [{\citenamefont {Astumian}\ and\ \citenamefont
  {Bier}(1994)}]{astumian_fluctuation_1994}%
  \BibitemOpen
  \bibfield  {author} {\bibinfo {author} {\bibfnamefont {R.~D.}\ \bibnamefont
  {Astumian}}\ and\ \bibinfo {author} {\bibfnamefont {M.}~\bibnamefont
  {Bier}},\ }\bibfield  {title} {\bibinfo {title} {Fluctuation driven ratchets:
  {Molecular} motors},\ }\href {https://doi.org/10.1103/PhysRevLett.72.1766}
  {\bibfield  {journal} {\bibinfo  {journal} {Physical Review Letters}\
  }\textbf {\bibinfo {volume} {72}},\ \bibinfo {pages} {1766} (\bibinfo {year}
  {1994})}\BibitemShut {NoStop}%
\bibitem [{\citenamefont {Sekimoto}(1997)}]{Sekimoto1997}%
  \BibitemOpen
  \bibfield  {author} {\bibinfo {author} {\bibfnamefont {K.}~\bibnamefont
  {Sekimoto}},\ }\bibfield  {title} {\bibinfo {title} {{Kinetic
  characterization of heat bath and the energetics of thermal ratchet
  models}},\ }\href {http://jpsj.ipap.jp/link?JPSJ/66/1234/} {\bibfield
  {journal} {\bibinfo  {journal} {Journal of the physical society of Japan}\
  }\textbf {\bibinfo {volume} {66}},\ \bibinfo {pages} {1234} (\bibinfo {year}
  {1997})}\BibitemShut {NoStop}%
\bibitem [{\citenamefont {Reimann}\ and\ \citenamefont
  {Hänggi}(2002)}]{reimann_introduction_2002}%
  \BibitemOpen
  \bibfield  {author} {\bibinfo {author} {\bibfnamefont {P.}~\bibnamefont
  {Reimann}}\ and\ \bibinfo {author} {\bibfnamefont {P.}~\bibnamefont
  {Hänggi}},\ }\bibfield  {title} {\bibinfo {title} {Introduction to the
  physics of {Brownian} motors},\ }\href
  {https://doi.org/10.1007/s003390201331} {\bibfield  {journal} {\bibinfo
  {journal} {Applied Physics A}\ }\textbf {\bibinfo {volume} {75}},\ \bibinfo
  {pages} {169} (\bibinfo {year} {2002})}\BibitemShut {NoStop}%
\bibitem [{\citenamefont {Astumian}\ and\ \citenamefont
  {Hänggi}(2002)}]{astumian_brownian_2002}%
  \BibitemOpen
  \bibfield  {author} {\bibinfo {author} {\bibfnamefont {R.~D.}\ \bibnamefont
  {Astumian}}\ and\ \bibinfo {author} {\bibfnamefont {P.}~\bibnamefont
  {Hänggi}},\ }\bibfield  {title} {\bibinfo {title} {Brownian {Motors}},\
  }\href {https://doi.org/10.1063/1.1535005} {\bibfield  {journal} {\bibinfo
  {journal} {Physics Today}\ }\textbf {\bibinfo {volume} {55}},\ \bibinfo
  {pages} {33} (\bibinfo {year} {2002})}\BibitemShut {NoStop}%
\bibitem [{\citenamefont {Reimann}(2002)}]{Reimann2002}%
  \BibitemOpen
  \bibfield  {author} {\bibinfo {author} {\bibfnamefont {P.}~\bibnamefont
  {Reimann}},\ }\bibfield  {title} {\bibinfo {title} {{Brownian motors: noisy
  transport far from equilibrium}},\ }\href
  {http://linkinghub.elsevier.com/retrieve/pii/S0370157301000813} {\bibfield
  {journal} {\bibinfo  {journal} {Physics Reports}\ }\textbf {\bibinfo {volume}
  {361}},\ \bibinfo {pages} {57} (\bibinfo {year} {2002})}\BibitemShut
  {NoStop}%
\bibitem [{\citenamefont {Feito}\ and\ \citenamefont
  {Cao}(2007{\natexlab{a}})}]{feito_time-delayed_2007}%
  \BibitemOpen
  \bibfield  {author} {\bibinfo {author} {\bibfnamefont {M.}~\bibnamefont
  {Feito}}\ and\ \bibinfo {author} {\bibfnamefont {F.~J.}\ \bibnamefont
  {Cao}},\ }\bibfield  {title} {\bibinfo {title} {Time-delayed feedback control
  of a flashing ratchet},\ }\href {https://doi.org/10.1103/PhysRevE.76.061113}
  {\bibfield  {journal} {\bibinfo  {journal} {Physical Review E}\ }\textbf
  {\bibinfo {volume} {76}},\ \bibinfo {pages} {061113} (\bibinfo {year}
  {2007}{\natexlab{a}})}\BibitemShut {NoStop}%
\bibitem [{\citenamefont {Craig}\ \emph {et~al.}(2008)\citenamefont {Craig},
  \citenamefont {Long}, \citenamefont {Parrondo},\ and\ \citenamefont
  {Linke}}]{craig_effect_2008}%
  \BibitemOpen
  \bibfield  {author} {\bibinfo {author} {\bibfnamefont {E.~M.}\ \bibnamefont
  {Craig}}, \bibinfo {author} {\bibfnamefont {B.~R.}\ \bibnamefont {Long}},
  \bibinfo {author} {\bibfnamefont {J.~M.~R.}\ \bibnamefont {Parrondo}},\ and\
  \bibinfo {author} {\bibfnamefont {H.}~\bibnamefont {Linke}},\ }\bibfield
  {title} {\bibinfo {title} {Effect of time delay on feedback control of a
  flashing ratchet},\ }\href {https://doi.org/10.1209/0295-5075/81/10002}
  {\bibfield  {journal} {\bibinfo  {journal} {Europhysics Letters (EPL)}\
  }\textbf {\bibinfo {volume} {81}},\ \bibinfo {pages} {10002} (\bibinfo {year}
  {2008})}\BibitemShut {NoStop}%
\bibitem [{\citenamefont {Roca}\ \emph {et~al.}(2014)\citenamefont {Roca},
  \citenamefont {Villaluenga},\ and\ \citenamefont
  {Dinis}}]{roca_optimal_2014}%
  \BibitemOpen
  \bibfield  {author} {\bibinfo {author} {\bibfnamefont {F.}~\bibnamefont
  {Roca}}, \bibinfo {author} {\bibfnamefont {J.~P.~G.}\ \bibnamefont
  {Villaluenga}},\ and\ \bibinfo {author} {\bibfnamefont {L.}~\bibnamefont
  {Dinis}},\ }\bibfield  {title} {\bibinfo {title} {Optimal protocol for a
  collective flashing ratchet},\ }\href
  {https://doi.org/10.1209/0295-5075/107/10006} {\bibfield  {journal} {\bibinfo
   {journal} {EPL (Europhysics Letters)}\ }\textbf {\bibinfo {volume} {107}},\
  \bibinfo {pages} {10006} (\bibinfo {year} {2014})}\BibitemShut {NoStop}%
\bibitem [{\citenamefont {Feito}\ and\ \citenamefont
  {Cao}(2007{\natexlab{b}})}]{Feito2007}%
  \BibitemOpen
  \bibfield  {author} {\bibinfo {author} {\bibfnamefont {M.}~\bibnamefont
  {Feito}}\ and\ \bibinfo {author} {\bibfnamefont {F.~J.}\ \bibnamefont
  {Cao}},\ }\bibfield  {title} {\bibinfo {title} {{Information and maximum
  power in a feedback controlled Brownian ratchet}},\ }\href
  {https://doi.org/10.1140/epjb/e2007-00255-7} {\bibfield  {journal} {\bibinfo
  {journal} {The European Physical Journal B}\ }\textbf {\bibinfo {volume}
  {59}},\ \bibinfo {pages} {63} (\bibinfo {year}
  {2007}{\natexlab{b}})}\BibitemShut {NoStop}%
\bibitem [{\citenamefont {Jarillo}\ \emph {et~al.}(2016)\citenamefont
  {Jarillo}, \citenamefont {Tangarife},\ and\ \citenamefont
  {Cao}}]{jarillo_efficiency_2016}%
  \BibitemOpen
  \bibfield  {author} {\bibinfo {author} {\bibfnamefont {J.}~\bibnamefont
  {Jarillo}}, \bibinfo {author} {\bibfnamefont {T.}~\bibnamefont {Tangarife}},\
  and\ \bibinfo {author} {\bibfnamefont {F.~J.}\ \bibnamefont {Cao}},\
  }\bibfield  {title} {\bibinfo {title} {Efficiency at maximum power of a
  discrete feedback ratchet},\ }\href
  {https://doi.org/10.1103/PhysRevE.93.012142} {\bibfield  {journal} {\bibinfo
  {journal} {Physical Review E}\ }\textbf {\bibinfo {volume} {93}},\ \bibinfo
  {pages} {012142} (\bibinfo {year} {2016})}\BibitemShut {NoStop}%
\bibitem [{NoteI()}]{NoteI}%
  \BibitemOpen
  \bibinfo {note} {Control memory positions $C_i$ with $i>k$ do not contribute
  to the entropy until they store a control action, as all or them were
  initially reset to a fixed value, e.g., $C_i=0$.}\BibitemShut {Stop}%
\bibitem [{NoteII()}]{NoteII}%
  \BibitemOpen
  \bibinfo {note} {Strictly speaking, we expect a time-periodic long-time
  behavior for the $x$ modulo $L$ problem, i.e., with periodic boundary
  conditions.}\BibitemShut {Stop}%
\bibitem [{NoteIII()}]{NoteIII}%
  \BibitemOpen
  \bibinfo {note} {For a stationary stochastic process, $H'_{\protect \text
  {s}}(M)$ is a monotonically decreasing function of $M$ with a well-defined
  limit for large $M$~\cite {cover_elements_2006}, this limit value equals
  $H'_{\protect \text {s}}(M_0)$.}\BibitemShut {Stop}%
\bibitem [{NoteIV()}]{NoteIV}%
  \BibitemOpen
  \bibinfo {note} {From a fundamental point of view, the fact that $M_{0}>1$
  means that $\protect \bm {\Gamma }^{M}$ is not a Markov process.}\BibitemShut
  {Stop}%
\bibitem [{NoteV()}]{NoteV}%
  \BibitemOpen
  \bibinfo {note} {The absolute maximum $k_B \ln 2$ may not be reachable for
  all sets of parameters, because these probabilities may not reach
  $1/2$.}\BibitemShut {Stop}%
\bibitem [{NoteVI()}]{NoteVI}%
  \BibitemOpen
  \bibinfo {note} {In the opposite limit, for $V_0\ll k_B T$, the effect of the
  confinement is negligible and the particle freely diffuses over the whole
  real line. Then, the maximum of $-\Delta \protect \tilde {S}_\protect \text
  {info}$ would be located very close to $a=1/2$ for all $\Delta
  t_m$.}\BibitemShut {Stop}%
\bibitem [{\citenamefont {He}\ \emph {et~al.}(2023)\citenamefont {He},
  \citenamefont {Cheong}, \citenamefont {Pradana},\ and\ \citenamefont
  {Chew}}]{He2023}%
  \BibitemOpen
  \bibfield  {author} {\bibinfo {author} {\bibfnamefont {L.}~\bibnamefont
  {He}}, \bibinfo {author} {\bibfnamefont {J.~W.}\ \bibnamefont {Cheong}},
  \bibinfo {author} {\bibfnamefont {A.}~\bibnamefont {Pradana}},\ and\ \bibinfo
  {author} {\bibfnamefont {L.~Y.}\ \bibnamefont {Chew}},\ }\bibfield  {title}
  {\bibinfo {title} {{Effects of correlation in an information ratchet with
  finite tape}},\ }\href {https://doi.org/10.1103/PhysRevE.107.024130}
  {\bibfield  {journal} {\bibinfo  {journal} {Physical Review E}\ }\textbf
  {\bibinfo {volume} {107}},\ \bibinfo {pages} {024130} (\bibinfo {year}
  {2023})}\BibitemShut {NoStop}%
\bibitem [{\citenamefont {Van~Kampen}(1992)}]{van_kampen_stochastic_1992}%
  \BibitemOpen
  \bibfield  {author} {\bibinfo {author} {\bibfnamefont {N.~G.}\ \bibnamefont
  {Van~Kampen}},\ }\href@noop {} {\emph {\bibinfo {title} {Stochastic processes
  in {Physics} and {Chemistry}}}}\ (\bibinfo  {publisher} {North-Holland},\
  \bibinfo {year} {1992})\BibitemShut {NoStop}%
\bibitem [{\citenamefont {Brey}\ and\ \citenamefont
  {Prados}(1993)}]{brey_normal_1993}%
  \BibitemOpen
  \bibfield  {author} {\bibinfo {author} {\bibfnamefont {J.~J.}\ \bibnamefont
  {Brey}}\ and\ \bibinfo {author} {\bibfnamefont {A.}~\bibnamefont {Prados}},\
  }\bibfield  {title} {\bibinfo {title} {Normal solutions for master equations
  with time-dependent transition rates: {Application} to heating processes},\
  }\href@noop {} {\bibfield  {journal} {\bibinfo  {journal} {Physical Review
  E}\ }\textbf {\bibinfo {volume} {47}},\ \bibinfo {pages} {1541} (\bibinfo
  {year} {1993})}\BibitemShut {NoStop}%
\bibitem [{\citenamefont {Marconi}\ \emph {et~al.}(2013)\citenamefont
  {Marconi}, \citenamefont {Puglisi},\ and\ \citenamefont
  {Vulpiani}}]{marconi_about_2013}%
  \BibitemOpen
  \bibfield  {author} {\bibinfo {author} {\bibfnamefont {U.~M.~B.}\
  \bibnamefont {Marconi}}, \bibinfo {author} {\bibfnamefont {A.}~\bibnamefont
  {Puglisi}},\ and\ \bibinfo {author} {\bibfnamefont {A.}~\bibnamefont
  {Vulpiani}},\ }\bibfield  {title} {\bibinfo {title} {About an {H}-theorem for
  systems with non-conservative interactions},\ }\href
  {https://doi.org/10.1088/1742-5468/2013/08/P08003} {\bibfield  {journal}
  {\bibinfo  {journal} {Journal of Statistical Mechanics: Theory and
  Experiment}\ ,\ \bibinfo {pages} {P08003}} (\bibinfo {year}
  {2013})}\BibitemShut {NoStop}%
\bibitem [{\citenamefont {García~de Soria}\ \emph {et~al.}(2015)\citenamefont
  {García~de Soria}, \citenamefont {Maynar}, \citenamefont {Mischler},
  \citenamefont {Mouhot}, \citenamefont {Rey},\ and\ \citenamefont
  {Trizac}}]{garcia_de_soria_towards_2015}%
  \BibitemOpen
  \bibfield  {author} {\bibinfo {author} {\bibfnamefont {M.~I.}\ \bibnamefont
  {García~de Soria}}, \bibinfo {author} {\bibfnamefont {P.}~\bibnamefont
  {Maynar}}, \bibinfo {author} {\bibfnamefont {S.}~\bibnamefont {Mischler}},
  \bibinfo {author} {\bibfnamefont {C.}~\bibnamefont {Mouhot}}, \bibinfo
  {author} {\bibfnamefont {T.}~\bibnamefont {Rey}},\ and\ \bibinfo {author}
  {\bibfnamefont {E.}~\bibnamefont {Trizac}},\ }\bibfield  {title} {\bibinfo
  {title} {Towards an {H}-theorem for granular gases},\ }\href
  {https://doi.org/10.1088/1742-5468/2015/11/P11009} {\bibfield  {journal}
  {\bibinfo  {journal} {Journal of Statistical Mechanics: Theory and
  Experiment}\ ,\ \bibinfo {pages} {P11009}} (\bibinfo {year}
  {2015})}\BibitemShut {NoStop}%
\bibitem [{\citenamefont {Plata}\ and\ \citenamefont
  {Prados}(2017)}]{plata_global_2017}%
  \BibitemOpen
  \bibfield  {author} {\bibinfo {author} {\bibfnamefont {C.~A.}\ \bibnamefont
  {Plata}}\ and\ \bibinfo {author} {\bibfnamefont {A.}~\bibnamefont {Prados}},\
  }\bibfield  {title} {\bibinfo {title} {Global stability and {H}-theorem in
  lattice models with nonconservative interactions},\ }\href
  {https://doi.org/10.1103/PhysRevE.95.052121} {\bibfield  {journal} {\bibinfo
  {journal} {Physical Review E}\ }\textbf {\bibinfo {volume} {95}},\ \bibinfo
  {pages} {052121} (\bibinfo {year} {2017})}\BibitemShut {NoStop}%
\bibitem [{\citenamefont {Oono}\ and\ \citenamefont
  {Paniconi}(1998)}]{oono_steady_1998}%
  \BibitemOpen
  \bibfield  {author} {\bibinfo {author} {\bibfnamefont {Y.}~\bibnamefont
  {Oono}}\ and\ \bibinfo {author} {\bibfnamefont {M.}~\bibnamefont
  {Paniconi}},\ }\bibfield  {title} {\bibinfo {title} {Steady {State}
  {Thermodynamics}},\ }\href@noop {} {\bibfield  {journal} {\bibinfo  {journal}
  {Progress of Theoretical Physics Supplement}\ }\textbf {\bibinfo {volume}
  {130}},\ \bibinfo {pages} {29} (\bibinfo {year} {1998})}\BibitemShut
  {NoStop}%
\bibitem [{\citenamefont {Hatano}\ and\ \citenamefont
  {Sasa}(2001)}]{hatano_steady-state_2001}%
  \BibitemOpen
  \bibfield  {author} {\bibinfo {author} {\bibfnamefont {T.}~\bibnamefont
  {Hatano}}\ and\ \bibinfo {author} {\bibfnamefont {S.-i.}\ \bibnamefont
  {Sasa}},\ }\bibfield  {title} {\bibinfo {title} {Steady-{State}
  {Thermodynamics} of {Langevin} {Systems}},\ }\href
  {https://doi.org/10.1103/PhysRevLett.86.3463} {\bibfield  {journal} {\bibinfo
   {journal} {Physical Review Letters}\ }\textbf {\bibinfo {volume} {86}},\
  \bibinfo {pages} {3463} (\bibinfo {year} {2001})}\BibitemShut {NoStop}%
\bibitem [{\citenamefont {Speck}\ and\ \citenamefont
  {Seifert}(2005)}]{speck_integral_2005}%
  \BibitemOpen
  \bibfield  {author} {\bibinfo {author} {\bibfnamefont {T.}~\bibnamefont
  {Speck}}\ and\ \bibinfo {author} {\bibfnamefont {U.}~\bibnamefont
  {Seifert}},\ }\bibfield  {title} {\bibinfo {title} {Integral fluctuation
  theorem for the housekeeping heat},\ }\href
  {https://doi.org/10.1088/0305-4470/38/34/L03} {\bibfield  {journal} {\bibinfo
   {journal} {Journal of Physics A: Mathematical and General}\ }\textbf
  {\bibinfo {volume} {38}},\ \bibinfo {pages} {L581} (\bibinfo {year}
  {2005})}\BibitemShut {NoStop}%
\bibitem [{\citenamefont {Speck}\ and\ \citenamefont
  {Seifert}(2006)}]{speck_restoring_2006}%
  \BibitemOpen
  \bibfield  {author} {\bibinfo {author} {\bibfnamefont {T.}~\bibnamefont
  {Speck}}\ and\ \bibinfo {author} {\bibfnamefont {U.}~\bibnamefont
  {Seifert}},\ }\bibfield  {title} {\bibinfo {title} {Restoring a
  fluctuation-dissipation theorem in a nonequilibrium steady state},\ }\href
  {https://doi.org/10.1209/epl/i2005-10549-4} {\bibfield  {journal} {\bibinfo
  {journal} {Europhysics Letters (EPL)}\ }\textbf {\bibinfo {volume} {74}},\
  \bibinfo {pages} {391} (\bibinfo {year} {2006})}\BibitemShut {NoStop}%
\bibitem [{\citenamefont {Van~den Broeck}\ and\ \citenamefont
  {Esposito}(2010)}]{van_den_broeck_three_2010}%
  \BibitemOpen
  \bibfield  {author} {\bibinfo {author} {\bibfnamefont {C.}~\bibnamefont
  {Van~den Broeck}}\ and\ \bibinfo {author} {\bibfnamefont {M.}~\bibnamefont
  {Esposito}},\ }\bibfield  {title} {\bibinfo {title} {Three faces of the
  second law. {II}. {Fokker}-{Planck} formulation},\ }\href
  {https://doi.org/10.1103/PhysRevE.82.011144} {\bibfield  {journal} {\bibinfo
  {journal} {Physical Review E}\ }\textbf {\bibinfo {volume} {82}},\ \bibinfo
  {pages} {011144} (\bibinfo {year} {2010})}\BibitemShut {NoStop}%
\bibitem [{\citenamefont {Feito}\ \emph {et~al.}(2009)\citenamefont {Feito},
  \citenamefont {Baltanás},\ and\ \citenamefont {Cao}}]{Feito2009a}%
  \BibitemOpen
  \bibfield  {author} {\bibinfo {author} {\bibfnamefont {M.}~\bibnamefont
  {Feito}}, \bibinfo {author} {\bibfnamefont {J.~P.}\ \bibnamefont
  {Baltanás}},\ and\ \bibinfo {author} {\bibfnamefont {F.~J.}\ \bibnamefont
  {Cao}},\ }\bibfield  {title} {\bibinfo {title} {{Rocking feedback-controlled
  ratchets}},\ }\href {https://doi.org/031128 10.1103/PhysRevE.80.031128}
  {\bibfield  {journal} {\bibinfo  {journal} {Physical Review E}\ }\textbf
  {\bibinfo {volume} {80}},\ \bibinfo {pages} {031128} (\bibinfo {year}
  {2009})}\BibitemShut {NoStop}%
\bibitem [{\citenamefont {Jarillo}\ \emph {et~al.}(2018)\citenamefont
  {Jarillo}, \citenamefont {Villaluenga},\ and\ \citenamefont
  {Cao}}]{Jarillo2018b}%
  \BibitemOpen
  \bibfield  {author} {\bibinfo {author} {\bibfnamefont {J.}~\bibnamefont
  {Jarillo}}, \bibinfo {author} {\bibfnamefont {J.~P.}\ \bibnamefont
  {Villaluenga}},\ and\ \bibinfo {author} {\bibfnamefont {F.~J.}\ \bibnamefont
  {Cao}},\ }\bibfield  {title} {\bibinfo {title} {{Reliability of rectified
  transport: Coherence and reproducibility of transport by open-loop and
  feedback-controlled Brownian ratchets}},\ }\href
  {https://doi.org/10.1103/PhysRevE.98.032101} {\bibfield  {journal} {\bibinfo
  {journal} {Physical Review E}\ }\textbf {\bibinfo {volume} {98}},\ \bibinfo
  {pages} {1} (\bibinfo {year} {2018})}\BibitemShut {NoStop}%
\bibitem [{\citenamefont {Pontryagin}(1987)}]{pontryagin_mathematical_1987}%
  \BibitemOpen
  \bibfield  {author} {\bibinfo {author} {\bibfnamefont {L.~S.}\ \bibnamefont
  {Pontryagin}},\ }\href@noop {} {\emph {\bibinfo {title} {Mathematical
  {Theory} of {Optimal} {Processes}}}}\ (\bibinfo  {publisher} {CRC Press},\
  \bibinfo {year} {1987})\BibitemShut {NoStop}%
\bibitem [{\citenamefont {Guéry-Odelin}\ \emph {et~al.}(2023)\citenamefont
  {Guéry-Odelin}, \citenamefont {Jarzynski}, \citenamefont {Plata},
  \citenamefont {Prados},\ and\ \citenamefont
  {Trizac}}]{guery-odelin_driving_2023}%
  \BibitemOpen
  \bibfield  {author} {\bibinfo {author} {\bibfnamefont {D.}~\bibnamefont
  {Guéry-Odelin}}, \bibinfo {author} {\bibfnamefont {C.}~\bibnamefont
  {Jarzynski}}, \bibinfo {author} {\bibfnamefont {C.~A.}\ \bibnamefont
  {Plata}}, \bibinfo {author} {\bibfnamefont {A.}~\bibnamefont {Prados}},\ and\
  \bibinfo {author} {\bibfnamefont {E.}~\bibnamefont {Trizac}},\ }\bibfield
  {title} {\bibinfo {title} {Driving rapidly while remaining in control:
  classical shortcuts from {Hamiltonian} to stochastic dynamics},\ }\href
  {https://doi.org/10.1088/1361-6633/acacad} {\bibfield  {journal} {\bibinfo
  {journal} {Reports on Progress in Physics}\ }\textbf {\bibinfo {volume}
  {86}},\ \bibinfo {pages} {035902} (\bibinfo {year} {2023})}\BibitemShut
  {NoStop}%
\bibitem [{\citenamefont {Chennakesavalu}\ and\ \citenamefont
  {Rotskoff}(2023)}]{chennakesavalu_unified_2023}%
  \BibitemOpen
  \bibfield  {author} {\bibinfo {author} {\bibfnamefont {S.}~\bibnamefont
  {Chennakesavalu}}\ and\ \bibinfo {author} {\bibfnamefont {G.~M.}\
  \bibnamefont {Rotskoff}},\ }\bibfield  {title} {\bibinfo {title} {Unified,
  {Geometric} {Framework} for {Nonequilibrium} {Protocol} {Optimization}},\
  }\href {https://doi.org/10.1103/PhysRevLett.130.107101} {\bibfield  {journal}
  {\bibinfo  {journal} {Physical Review Letters}\ }\textbf {\bibinfo {volume}
  {130}},\ \bibinfo {pages} {107101} (\bibinfo {year} {2023})}\BibitemShut
  {NoStop}%
\bibitem [{\citenamefont {Blaber}\ and\ \citenamefont
  {Sivak}(2022)}]{blaber_optimal_2022}%
  \BibitemOpen
  \bibfield  {author} {\bibinfo {author} {\bibfnamefont {S.}~\bibnamefont
  {Blaber}}\ and\ \bibinfo {author} {\bibfnamefont {D.~A.}\ \bibnamefont
  {Sivak}},\ }\href {http://arxiv.org/abs/2212.00706} {\bibinfo {title}
  {Optimal {Control} in {Stochastic} {Thermodynamics}}} (\bibinfo {year}
  {2022}),\ \bibinfo {note} {arXiv:2212.00706 [cond-mat]}\BibitemShut {NoStop}%
\bibitem [{\citenamefont {Ehrich}\ \emph {et~al.}(2022)\citenamefont {Ehrich},
  \citenamefont {Still},\ and\ \citenamefont {Sivak}}]{Ehrich2022}%
  \BibitemOpen
  \bibfield  {author} {\bibinfo {author} {\bibfnamefont {J.}~\bibnamefont
  {Ehrich}}, \bibinfo {author} {\bibfnamefont {S.}~\bibnamefont {Still}},\ and\
  \bibinfo {author} {\bibfnamefont {D.~A.}\ \bibnamefont {Sivak}},\ }\href
  {https://doi.org/10.48550/arxiv.2206.10793} {\bibinfo {title} {{Energetic
  cost of feedback control}}} (\bibinfo {year} {2022}),\ \Eprint
  {https://arxiv.org/abs/2206.10793} {arXiv:2206.10793} \BibitemShut {NoStop}%
\bibitem [{\citenamefont {VanSaders}\ and\ \citenamefont
  {Vitelli}(2023)}]{VanSaders2023}%
  \BibitemOpen
  \bibfield  {author} {\bibinfo {author} {\bibfnamefont {B.}~\bibnamefont
  {VanSaders}}\ and\ \bibinfo {author} {\bibfnamefont {V.}~\bibnamefont
  {Vitelli}},\ }\href {https://doi.org/10.48550/arxiv.2302.07402} {\bibinfo
  {title} {{Informational active matter}}} (\bibinfo {year} {2023}),\ \Eprint
  {https://arxiv.org/abs/2302.07402} {arXiv:2302.07402} \BibitemShut {NoStop}%
\bibitem [{NoteVII()}]{NoteVII}%
  \BibitemOpen
  \bibinfo {note} {Intuitively, this means that there is no loss of
  information, which hints at the continuity of the entropy that is proved here
  mathematically.}\BibitemShut {Stop}%
\end{thebibliography}

\end{document}